# Know Their Name: An Anthology of Great Women in the Shadows

Janan Arslan[1*], Sepinoud Azimi[2], Lina Sami[1], Farah Ajili[3], Daniela Domingues[1], Deniz Akpinar[4], Lærke Vinther Christiansen[2], Violetta Zujovic[1], and Kurt K. Benke[5]

1. *Sorbonne Université, Institut du Cerveau—Paris Brain Institute—ICM, CNRS, Inria, Inserm, AP-HP, Hopital de la Pitié Salpetriere, Paris, France*

2. *Technology, Policy and Management, Delft University of Technology, Delft, The Netherlands*

3. *Maison des Artistes, Montpellier, France*

4. *Erzincan Binali Yıldırım University, Faculty of Education Department of Social Sciences Education, Erzincan, Turkiye*

5. *School of Engineering, University of Melbourne, Parkville, Victoria, Australia*

*Corresponding author, e-mail: janan.arslan@icm-institute.org.

## Abstract

There has been a long history of women innovators producing outstanding contributions to society and public benefit yet having their work passed over or sidelined or attributed to male colleagues. This phenomenon has been coined the "Matilda Effect". The amendments to the record of human achievements are now taking place, with an increasing pace in recent times due to greater social enlightenment and awareness and the interest in social justice. However, there remains a gap that must be addressed.  In this article, we demonstrate the disparity in scientific recognition for a handful of case studies through search data collected via Google Trends and plotted as time-series figures and choropleth maps. Search trends reflect a noticeable divergence between recognition of female and male innovators. However, we note that in more well-known cases of the Matilda Effect, such as the historical account of Rosalind Franklin vs. James Watson and Francis Crick, the differences become less pronounced, emphasizing the importance of publicizing recognition. In response to this revelation, this article presents the stories of several women innovators and their great achievements. We identify the truth behind several discoveries and inventions, while revealing the full nature of this historical problem of social exclusion.

# Introduction

*"No assertion in reference to woman is more common than that she possesses no inventive or mechanical genius, even the United States census failing to enumerate her among the inventors of the country. ………"* – Matilda Joslyn Gage, Woman as an Inventor [1]

The "Matilda Effect" is a generic term for bias against contributions from female scientists with their achievements credited to male colleagues. The most famous case is probably that of Rosalind Franklin and her contribution to the discovery of DNA but ignored in the award of the Nobel Prize.

The Matilda Effect was first articulated by the suffragette Matilda Joslyn Gage. However, the definition did not receive its official name "Matilda" until 1993, when it was coined by historian Margaret W. Rossiter to honor of the late Gage [2]. Prior to its official naming, another like term had surfaced known as the "Matthew Effect", made famous by Robert K. Merton in 1968 [3]. The Matthew Effect refers to how members of a group receive disproportionate recognition of a piece of work, in which those of higher ranking or renown will be given most if not all credit of said work over those who do not hold a high enough ranking to be acknowledged, even if the majority of the work was a result of their efforts. Merton's original choice of "Matthew" for the name stems from the second half Matthew 13:12 in the Bible, which reads *"Whoever has will be given more, and they will have an abundance. Whoever does not have, even what they have will be taken from them."* The latter sentence underpins the lack of recognition for those who have little to begin with, and in many sense reflects how absolute power can sway truth and balance [2]. The Matilda Effect is a continuation of the Matthew Effect, however it specifically focuses on how the contributions made by women are often overlooked or credited to men [4].

The Matthew and Matilda Effects are ever so present in academia. For example, a study conducted by Patel *et al.* (2021) assessed the Matilda Effect for published award recipients in the field of hematology and oncology over two-three decades. The study found that out of a total of 1,642 awardees, 77.9% were men [5]. Historically, we know many female innovators have been overlooked. That said, some cases are more well-known than others. Rosalind Franklin is the prime example that comes to mind. Franklin – an accomplished X-ray crystallographer, chemist, and molecular biologist – had demonstrated the presence of the helical structure of DNA using X-ray photography. However, the discovery of this was credited to James Watson and Francis Crick; anyone who has studied biology will recall seeing these names in their textbooks when learning about DNA. This is because Maurice Wilkins secretly copied the groundbreaking work of Franklin and passed this information along to Franklin's rival, Watson. While the world is now slowly catching up to the this truth, Franklin's legacy had suffered as it was Watson and Crick who were given an unprecedented recognition that should have gone to Franklin [6].

Unfortunately, Franklin's story is not an isolated incident, although thankfully one which has gained traction over the years with the truth prevailing. There are many stories akin to Franklin's that are also deserving of being told.

## The Recognition Gap

To demonstrate the disparity in recognition, we begin this article by illustrating a small case study which showed the interest in female innovators as compared to their male colleagues. Collecting data from Google Trends, we conducted three searches for illustrative purposes: (i) Marthe Gautier vs. Jérôme Lejeune, (ii) Marianne Weber vs. Max Weber, and (iii) Flora Tristan vs. Karl Marx and Friedrich Engels. Searches were investigated from the earliest possible data available (2004) until the time of the writing of this publication (July 2024) [1/1/04 - 7/28/24], thus trends reflect interests in these individuals over the span of 20 years. All data collected and code for analyses can be found in **Supplementary Information**. Observations were made for trends over time as well as trends on a global scale (i.e., the prevalence of searches in different regions) using the statistical software R (v. 4.2.0) and packages such **sf**, **ggplot2**, and **rnaturalearth** [7-9]. The objective of this case study was to illustrate how, despite the efforts of these female innovators, their male colleagues are the ones still getting the recognition. **Figures 1-3** illustrate search trends over time, while **Figures 4-6** illustrate choropleth maps that demonstrate global prevalence. We can see immediately in these figures that the global recognition and search interests for the male counterparts far exceeds that of the female counterparts.

Noting this disparity, we then proceeded to re-run the same data collection and visualization process with a focus on the more well-known case of Rosalind Franklin vs. James Watson and Francis Crick. As can be seen in **Figures 7** and **8**, the gap in recognition is far smaller as compared to our initial three case studies. There is a particular peak in the time trends which can be found in **Figure 7** against Franklin's name. This peak correlates with searches conducted on July 2013, and is associated with a tribute made by Google on 24th July 2013 themselves to honor the legacy of this (almost) forgotten legend [10]. This further iterates the need to publicize recognition, bringing these heroines into the light so their worldly contributions too can be recognized.

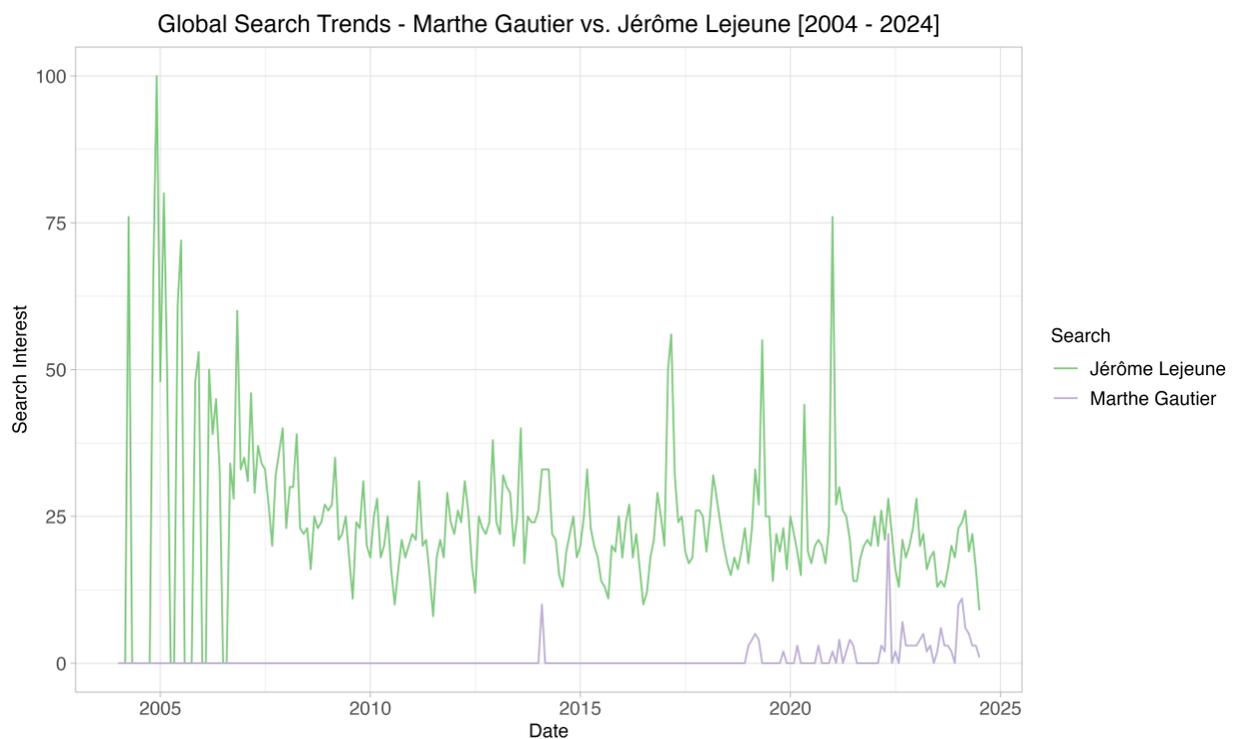

**Figure 1. Global Search Trends in Google for Marthe Gautier vs. Jérôme Lejeune [2004 - 2024].** Search trends for Jérôme Lejeune (green) far outweigh that of Marthe Gautier (purple) – a French doctor who discovered the Down's syndrome was characterized by the presence of an additional chromosome; a discovery for which Lejeune took credit for. Gautier was eventually credited as the discoverer by Inserm - the French national research organization.

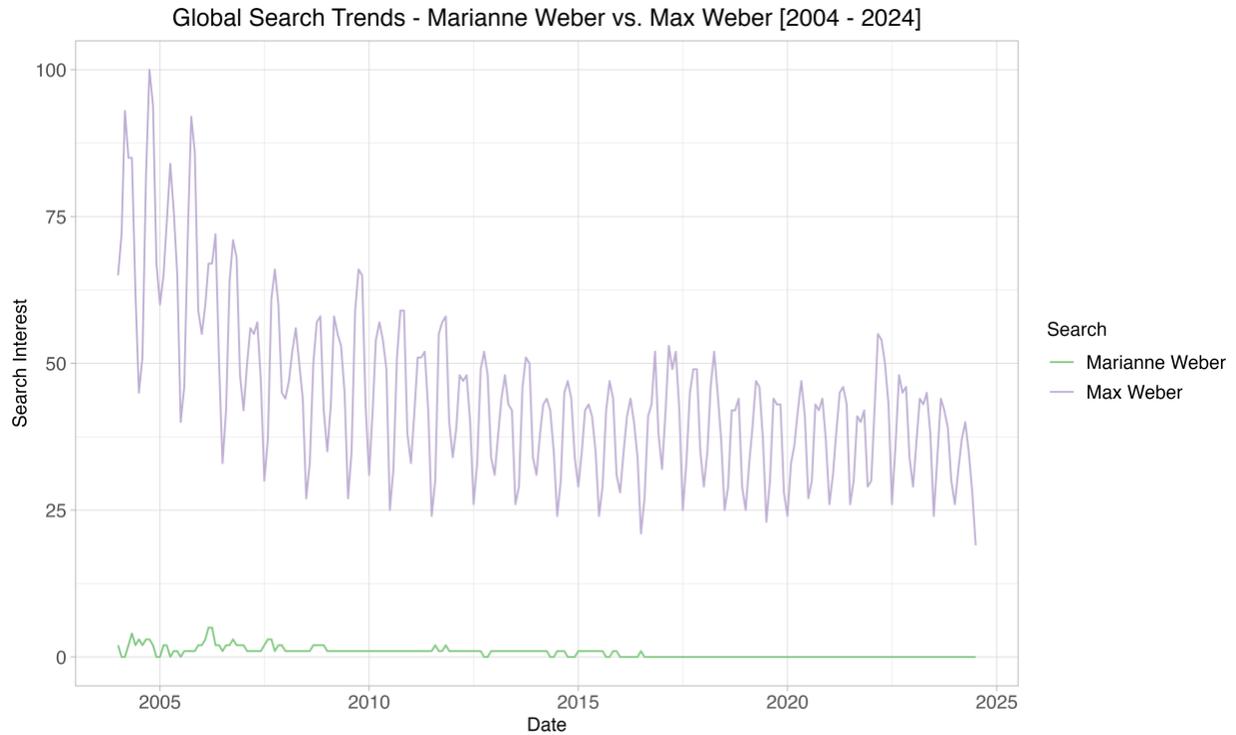

**Figure 2. Global Search Trends in Google for Marianne Weber vs. Max Weber [2004 - 2024].** Marianne (green) and Max Weber (purple) were both German sociologists. However, Marianne's legacy was overshadowed by that of her husband's, with her often being referred to as the wife (and later the widow) of the famous Max Weber. These search trends solidify she has been denied the rightful recognition she deserves.

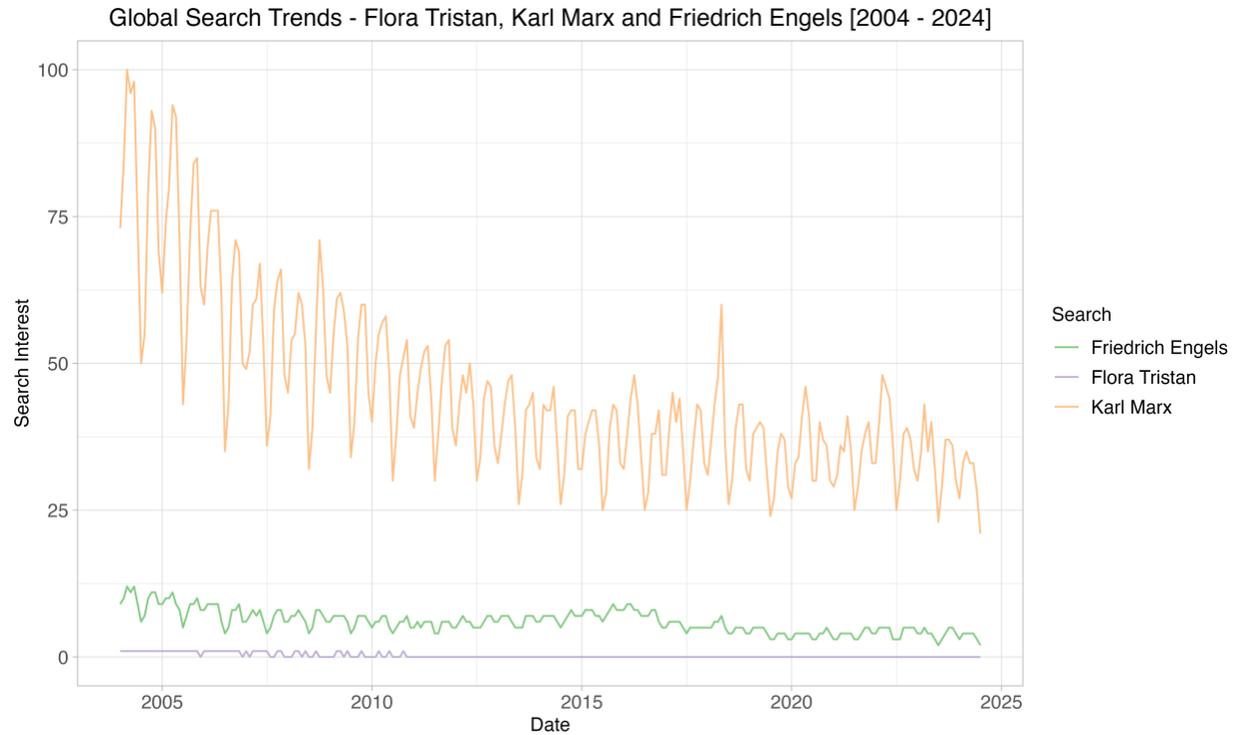

**Figure 3. Global Search Trends in Google for Flora Tristan vs. Friedrich Engels and Karl Marx.** Although all philosophers and social scientists, and despite Tristan's *The Workers' Union* being published five years before Marx and Engels *The Communist Manifesto*, the search trends favor Marx (orange), followed by Engels (green). We see Tristan's (purple) search queries flatlining around the early 2010s.

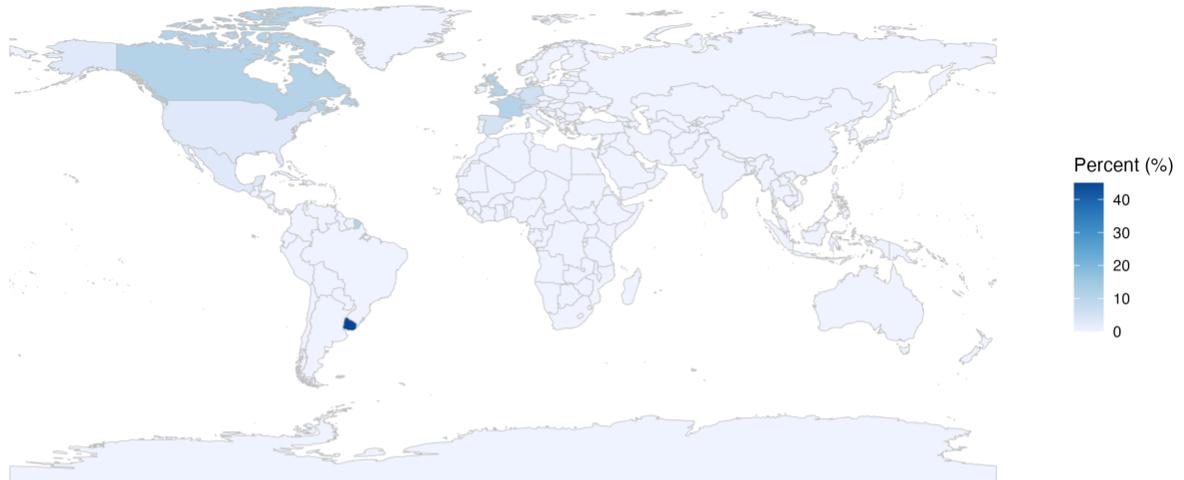

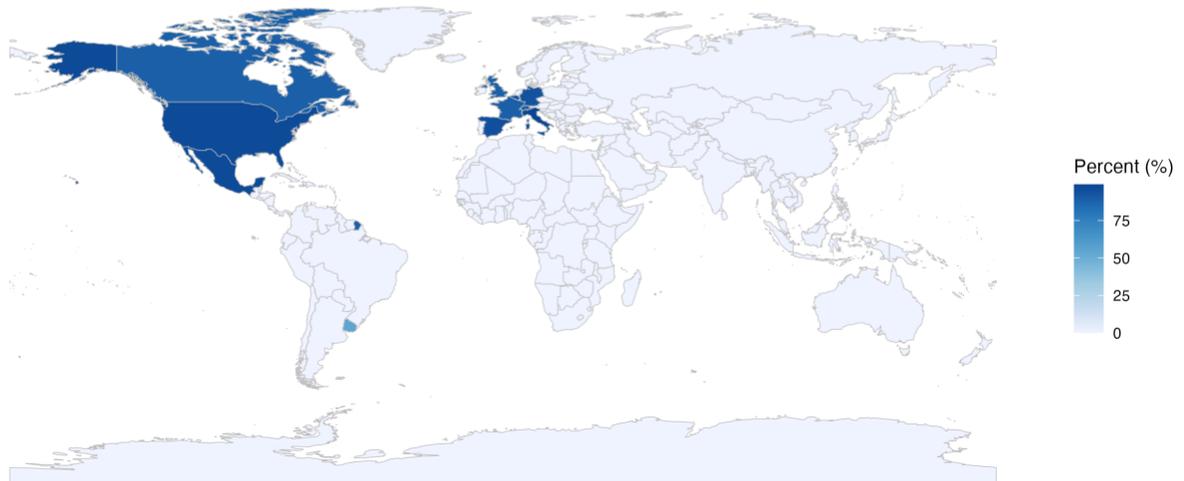

**Figure 4. Choropleth Maps using Google search data for Marthe Gautier vs. Jérôme Lejeune [2004-2024].** With a marginal exception of Uruguay (in which 45% of searches were for Gautier and 55% for Lejeune), most search activities across different countries, such as France (the native country of both individuals), Canada, Mexico, etc. predominantly favored Lejeune. The division of search queries between the two names resulted in a search prevalence ranging from 55 – 99% for Lejeune and 1 – 45 % for Gautier.

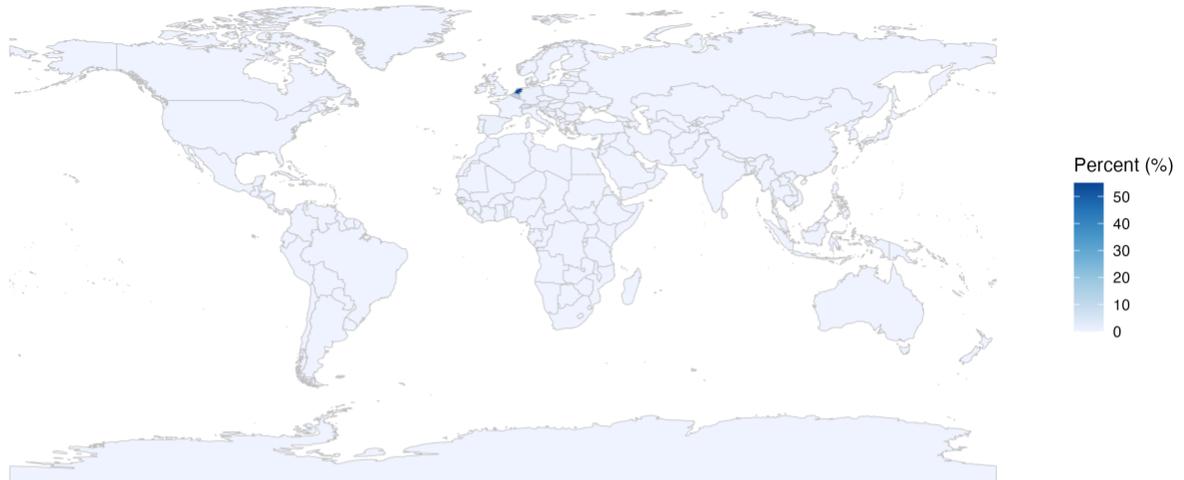

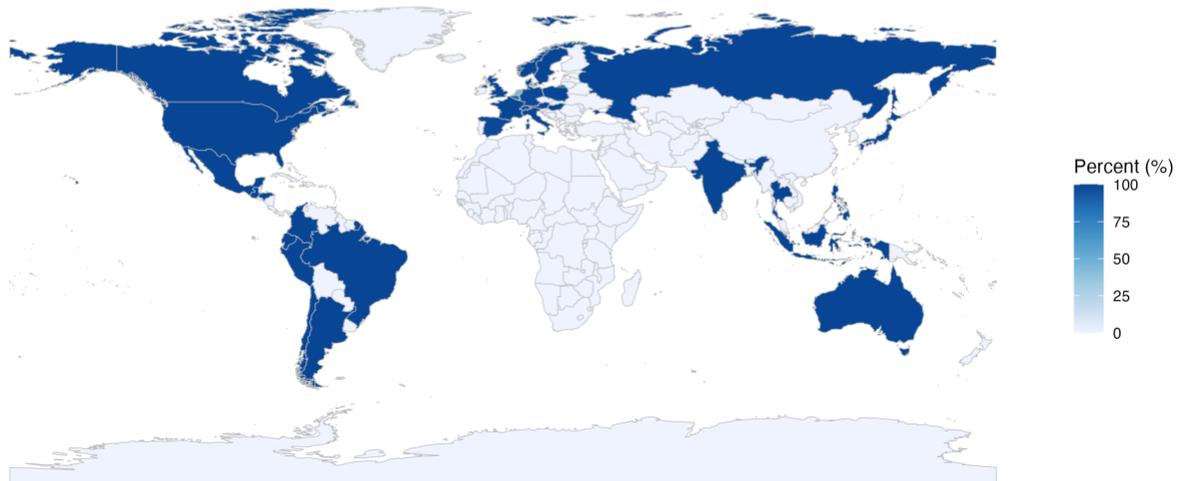

**Figure 5. Choropleth Maps using Google search data for Marianne Weber vs. Max Weber [2004-2024].** With the exception of the Netherlands (in which 55% of searches were for Marianne and 45% for Max), most search activities across different countries, such as Germany (the native country of both individuals), Belgium, Spain, India, Japan, etc. predominantly favored Max. The division of search queries between the two names resulted in a search prevalence ranging from 45 – 100% for Max and 0 – 55 % for Marianne.

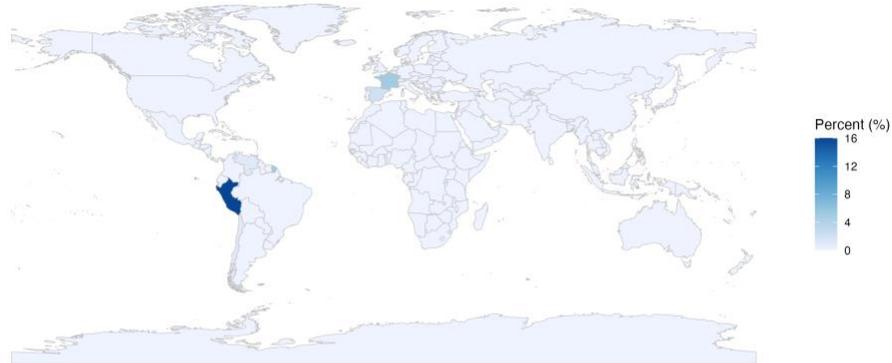

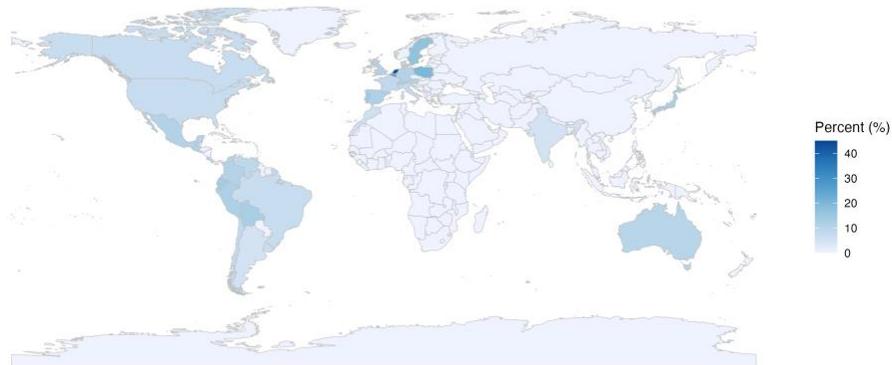

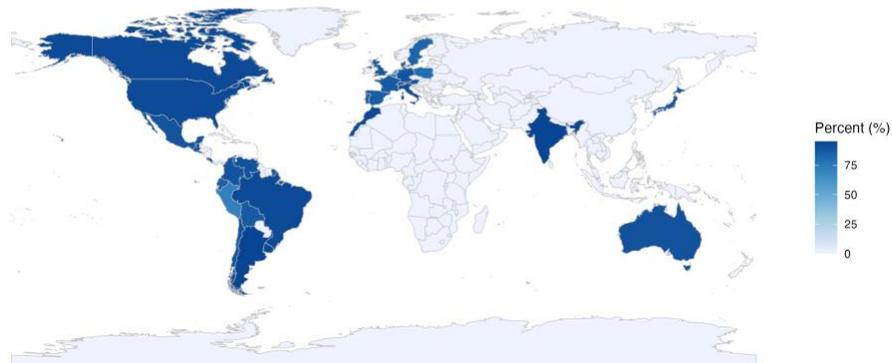

**Figure 6. Choropleth Maps using Google search data for Flora Tristan vs. Friedrich Engels, and Karl Marx [2004-2024].** Marx-related searches dominated this analysis. Even in France and Peru, the French-Peruvian Tristan had her 'largest' search queries sitting at 16%, while 72% of the searches were related to Marx, and the remaining 12% with Engels. Search queries across different nations ranged from 0 – 16% for Tristan, 5 – 45 % for Engels, and 55 – 95% for Marx.

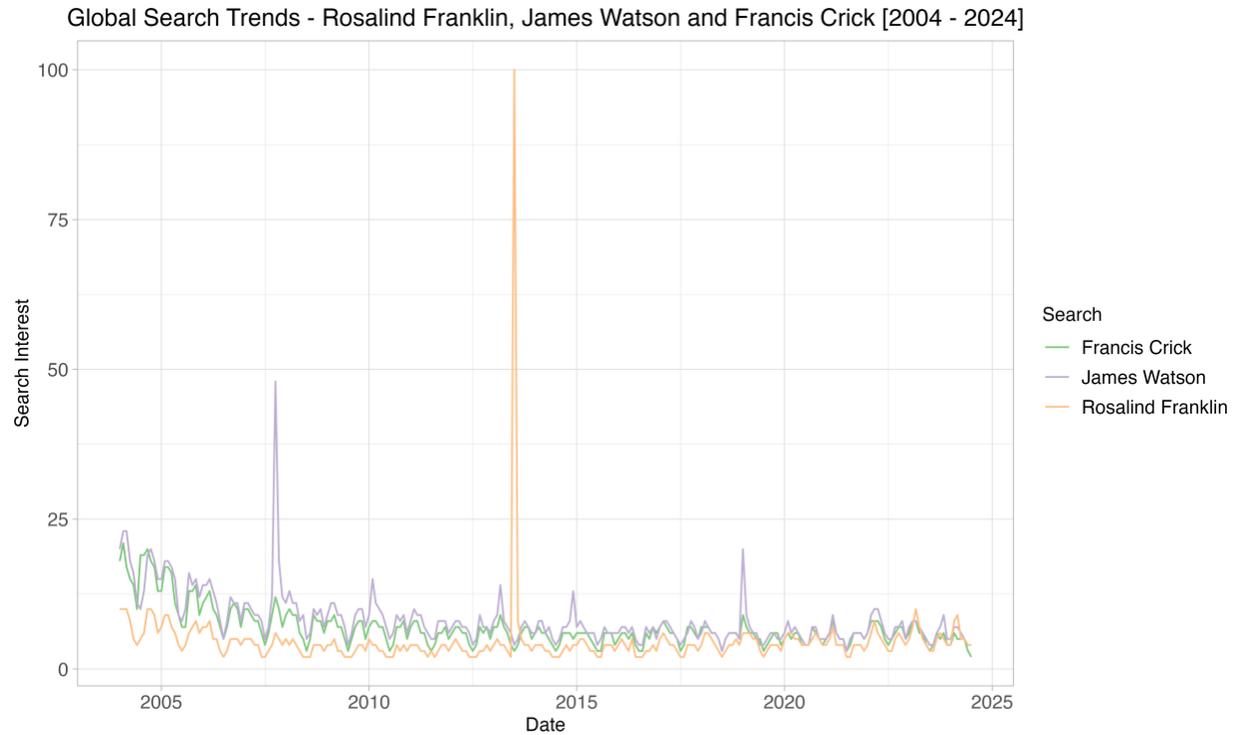

**Figure 7. Global Search Trends in Google for Rosalind Franklin vs. James Watson and Francis Crick.**

Franklin's story is one that has gained the most traction over time. This is evident as search trends for Franklin (orange) are closer in volume and distribution to that of Watson (purple) and Crick (green). This plot demonstrates the importance of publicizing recognition. The peak for Franklin correlates with a tribute made by Google on 24[th] July 2013, further iterating that interests for female scientists can significantly increase with the appropriate exposure.

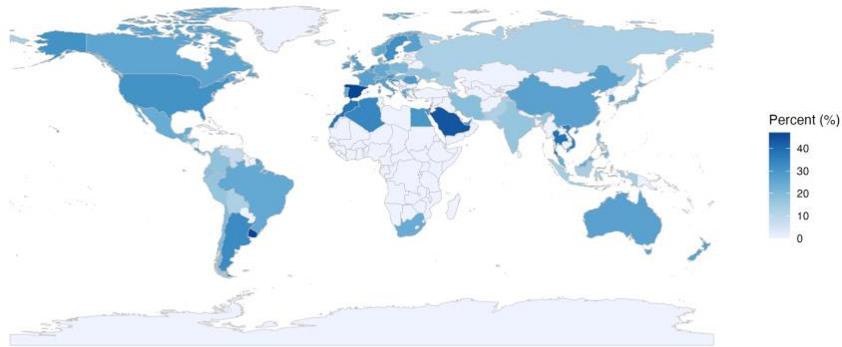

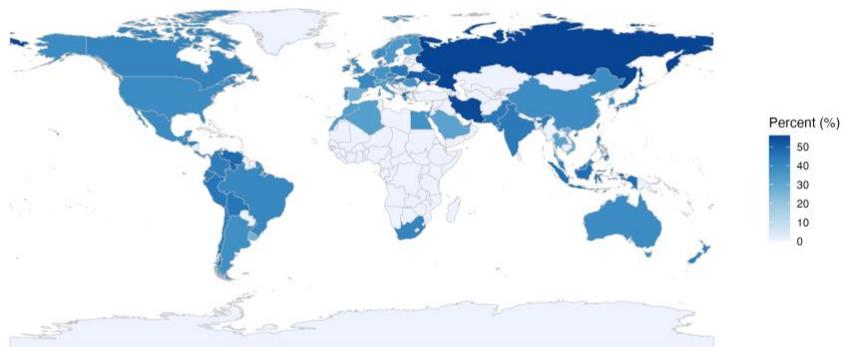

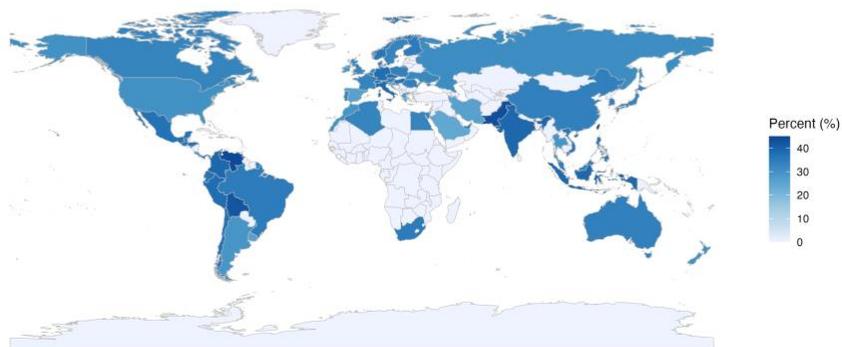

**Figure 8. Choropleth Maps using Google search data for Rosalind Franklin, James Watson, and Francis Crick [2004-2024].** While not perfect, results for Franklin are more favorable as compared to previous case studies. For example, in some countries, search queries for Franklin are higher than for Watson and Crick. In Uruguay and Spain, for example, Franklin had 47% of the search queries, Watson 27% and Crick 26%. Most countries appear to equally or near-equally search for the three scientists (e.g., Croatia, Sweden, Austria, Algeria, etc.), while some still predominantly search for Watson and Crick over Franklin (e.g., India, Bolivia, Peru, etc.).

# Turning the Tide

The *Know Their Name (KTN)* series highlights the gap in recognition between male and female scientists. There is a need to describe the struggles of innovative women who have (almost) disappeared from the pages of recorded history. The narratives presented in this article are based on the collective knowledge of the authors. However, there are undoubtedly many other cases in need of investigation, and it is our hope that the current paper will serve as a catalyst for further study and publication. We invite readers to join our KTN collective and contribute to future articles in which we continue to tell the stories of those whose recognition is long overdue.

The current paper features Hedy Lamarr, Marthe Gautier, Kathleen Lonsdale, Marguerite Perey, Cecilia Payne, Alice Ball, Emmy Noether, Flora Tristan, Jocelyn Bell Burnell, Marianne Weber, Chien-Shiung Wu, and Elinor Claire Ostrom. This inaugural edition of KTN further honors the charitable, humanitarian, and underappreciated work by Anatolian women who established Darüşşifas (Arabic for 'hospitals', with its literal translation being 'healing/health homes'). Finally, while honoring the past, we celebrate modern female STEAM leaders today through our Section titled "The Next Generation". In this first edition, author Lina Sami interviewed and wrote a piece regarding Farida Fassi – a Moroccan researcher and professor of Physics.

# Hedy Lamarr

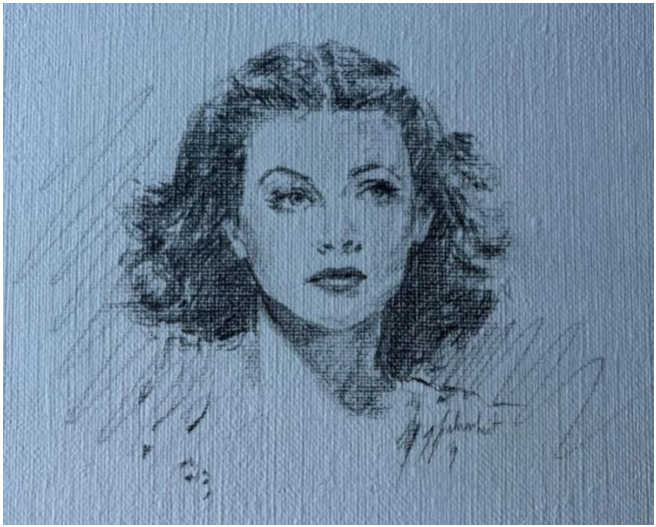

**Figure 9. Hedy Lamarr.** Famed for her beauty, Lamarr's technology talents and genius were as awe-inspiring as her presence on the silver screen, as is evidenced by her co-invention of the frequency-hoping system – a technology that underpins most wireless communications systems today. [*Illustrated by Farah Ajili, contributing author, 2024*]

*"My face has been my misfortune…A mask I cannot remove. I must live with it. I curse it."* – Hedy Lamarr.

Hedy Lamarr is perhaps a surprising example of an innovator in science and technology who was overlooked in the pages of history (**Figure 9**). This is largely because she achieved her initial fame in the Performing Arts rather than Technology. Hedy Lamarr, born Hedwig Eva Maria Kiesler in 1914 in Austria, was a celebrated Hollywood actress during its Golden Age in the 1940s and 1950s. What was less known and only has recently surfaced is Lamarr's ingenuity and co-invention of the frequency-hopping spread spectrum system – an ingenious concept that underpins many contemporary wireless communication systems, including Wi-Fi, Bluetooth, and radio technologies such as the code-division multiple access (CDMA).

Lamarr was famed for her beauty, as evident by the following quotes (which constitute only a handful of the poetic descriptions of Lamarr). It has even been suggested Lamarr was the inspiration behind Bob Kane's creation *CatWoman* [11].

> *"Miss Lamarr doesn't have to say 'Yes'. All she has to do is yawn…In her perfect will-lessness, Miss Lamarr is, indeed, identified metaphysically with her mesmeric midnight captor, the loving male."* - Park Tyler [12].

> *"One could feel the audience's anticipation of seeing her face for the first time. It was palpable. Sitting there in the dark, when the shadowed image of Hedy Lamarr suddenly turns her face full to the camera, the impact was audible*." – Billie Melba Fuller describing the audiences' captivation when Lamarr first graced the silver screen in her debut film *Algiers* [13].

One could feel the awe laced within these words in their description of the Hollywood starlet. There is no doubt that Lamarr was indeed beautiful. However, this beauty represented a single aspect of a wonderfully multifaceted character. In fact, Lamarr's beauty and star quality were so bright, it blinded onlookers to her other, most prominent characteristics: her mind, independence, creativity, and sheer willfulness coupled with determination. Behind the long lashes, green eyes, and porcelain skin that gained her notoriety in tinsel town, there was a science enthusiast and an innovative genius with a penchant for mathematics and engineering. She was multidisciplinary and displayed all the qualities of a polymath.

Lamarr's first love was certainly acting, although her interest in technology can be traced back to her childhood conversations with her father Emil, who would explain to his bright young daughter how the world would work. He encouraged her to formulate her own independence and critical thinking from a young age (for example, a 5-year old Lamarr would dismantle and reassemble her music box to understand how it worked) [14, 15]. Her fierce independence became apparent at the tender age of 16 years when she decided to forge a handwritten letter from her mother to skip school one day with the intention of securing a job at Austria's largest film studio at the time - Sascha-Film. Managing to secure a job as a script clerk, Lamarr would toil away while keeping her eyes peeled for potential opportunities within the star-studded walls. One day, she eavesdropped on a conversation between directors with regards to an acting job. Taking this as her cue, Lamarr immediately went to freshen her make-up before boldly approaching one of the assistant directors and announcing that *she* was *perfect* for the role of the secretary in the upcoming film. Her daredevil and charismatic nature of course landed her the small part in the film *Gold on the Street* which was released in 1930 [16].

This love of acting was momentarily swept under the rug during her first and tumultuous marriage to Friedrich "Fritz" Mandl who prevented her from pursuing her career. One advantage of this union for Lamarr was Mandl's business. Making a living from military arms and munitions, Mandl would bring Lamarr to his business meetings, where she gained incredible tech know-how as related to warfare. She'd intently listen in on the conversations surrounding her, while playing her most dreaded role as the dutiful wife to an unbearable husband. Eventually tiring from the frustration of dealing with Mandl's overbearing and controlling ways, Lamarr decided to put her individuality, independence, and career ahead of her stifling marriage. She first ended up finding herself in Paris then in London. In London she met Louis B. Mayer – co-founder of Metro-Goldwyn-Mayer (MGM) studios. He promoted her as "the world's most beautiful woman", and eventually "loaned" her to take part in the film *Algiers*. Over time, Lamarr found herself typecast as the beautiful seductress. Despite the fame she found on the silver screens, playing the femme fatale repeatedly bored Lamarr [17]. The novelty of the limelight wore thin, and Lamarr found herself seeking more challenging endeavors to stimulate her bright mind.

Enter stage left George Antheil. Antheil was a pianist and avant-garde composer, known for the *Ballet Mécanique* (1924) [18]. Lamarr met Antheil - her Hollywood neighbor - at a dinner party. The legend is that their initial discussions were around endocrinology. However, upon discovering Antheil's background as a weapons inspector, the conversations soon evolved into weaponry (naturally) [19]. Lamarr remembered back to her days in Europe with Mandl, recalling discussions around issues with guidance systems for remote-controlled torpedoes [15, 20]. There was a common concern about the relative ease of jamming or sending false remote-control signals, which could cause the torpedoes to go off course [21]. Previous torpedo guidance systems relied on a single, matched frequency between the torpedo and the guiding ship. This made it easy for an enemy to identify and interfere with the signal. Lamarr developed the idea of working within a bandwidth of multiple matching frequencies; if the enemy had discovered one frequency, it would have little consequence, as the message guiding the torpedo from the ship would be carried (and oscillate) between various frequencies in short bursts. Using his background in piano mechanisms, Lamarr and Antheil developed the frequency-hoping system, where a signal would hop among 88 frequencies – similar to that of the 88 keys on a piano keyboard [19].

Armed with their innovation, and joined by a University Professor in electrical engineering from Caltech, along with an attorney from Lyon & Lyon, they registered the patent "Secret Communication System" (US Patent 2,292,387) which was filed in 1941 and assigned to the US Navy [21]. However, a commanding officer in the Navy rejected the patent when they had discovered who was behind it [17, 20]. Lamarr, instead, provided for war efforts by using her star status to sell war bonds [17]. The true impact of their work wasn't acknowledged until the 1990s, long after others had built upon and profited from similar innovations without crediting or compensating Lamarr and Antheil. It was not until the 1990s that some recognition was given to the pair.

When awarded the Electronic Frontier Foundation award for her discovery, Lamarr rightfully remarked "It's about time" [19]. In 2014, Lamarr and Antheil were added to the National Inventors Hall of Fame [17].

# Marthe Gautier

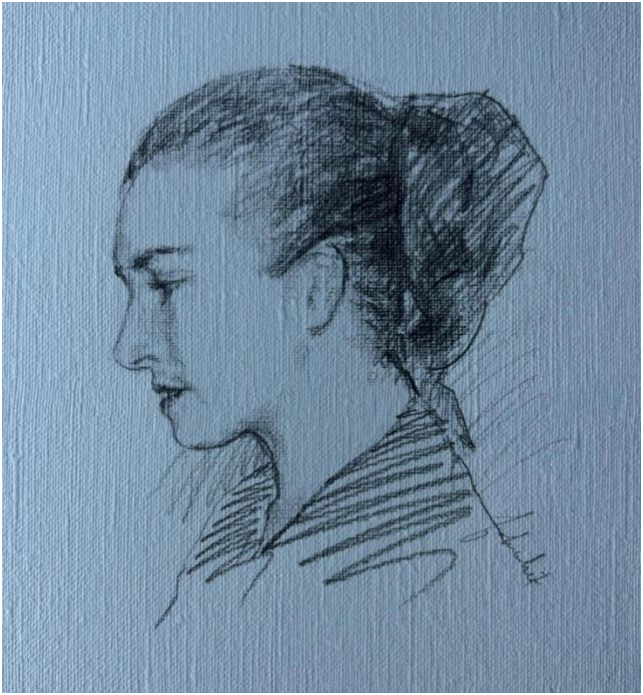

*"I am confident that, in the end, common sense and justice will prevail. I'm an optimist, brought up on the belief that if you wait to the end of the story, you get to see the good people live happily ever after."* – Cat Stevens

**Figure 10. Marthe Gautier.** Gautier discovered the additional chromosome in Down syndrome patients. However, for most of her scientific career, this discovery was credited to her colleague, Jérôme Lejeune, who borrowed her slides with the promise of taking better pictures. Lejeune did not keep his promise and instead took credit for her work. [*Illustrated by Farah Ajili, contributing author, 2024*]

Marthe Gautier (**Figure 10**) was a French doctor (pediatric cardiologist) and researcher who in 1958 discovered that Down's syndrome was characterized by an extra chromosome. However, her discovery was overlooked for half a Century when a male colleague, Jérôme Lejeune, borrowed her slide with the promise of taking better pictures of the chromosome for Gautier. The slide was never returned and the colleague then published a paper six months later with his name first and her name second [22]. There were even follow-up publications, one which particularly stands out because of its unfortunate title - *The Chromosomes of Man* [23]. However, this piece with its misogynistic title was simply a reply to a critic who was not convinced of the additional chromosome (the reply justified this skepticism as the result of a misunderstanding stemming from the use of the French word "supplémentaire"). No doubt, even such a reply would have felt like salt on an open wound for Gautier. Furthermore, it has been said that Gautier's name was incorrectly spelt initially as Gauthier. While later corrected, this oversight alone shows the blatant disregard for the discoverer [24].

Gautier came from a modest background from a farming family, being the fifth of seven children [25]. Born September 10[th], 1925, Gautier followed in the footsteps of her elder sister Paulette, who was studying medicine in Paris until her untimely demise by German troops in 1944. A piece of advice given to Gautier by her sister was: "If you're a woman, and you're not the boss's daughter, you have to be twice as good to succeed" [24]. Gautier remarked that from that point on, she had to achieve success for herself and for her late sister. This led her to completing her medical studies in 1950 and winning a prestigious medical internship to boot – an achievement with a huge gender imbalance at the time, given that out of the 80 recipients, only two were women. She eventually found herself in Harvard in September 1955 with a 1-year visa under her belt. During her time at Harvard, she learnt new techniques for treating rheumatic fever and heart disease in children. In parallel, she held a position as a technician in a cell-culture lab and learnt how to grow human cells in a dish [24, 25].

Upon returning back to Paris, Gautier did not receive the post she had been promised before her departure. She instead took up a poorly paid position with Raymond Turpin, the head of Hôpital Trousseau, who she had no prior association with. Following a Copenhagen conference in 1956, Turpin announced that the number of chromosomes in humans were 46 and not 48 as had been previously assumed. But, following this announcement, he stated there was no one in France who could do cell cultures to count the number of chromosomes in Down syndrome. Taken aback by the remark, Gautier decided to use her experience in the United States, responding with, "If you want, I'll do it, if someone lends me a laboratory". After finally acquiring tissues from children with Down syndrome, Gautier noticed an unmistakable difference: affected children consistently had a chromosomal count of 47, which differed from the 46 noted in the control samples. However, the chromosome was small, and she did not have a photomicroscope – a microscope with a light source and camera to produce a photograph. Thus, Gautier was not able to validate her finding. She handed her slides over to Lejeune, who took them with the promise of photographing them at a laboratory with better equipment. Lejeune took the photographs, but never showed them to Gautier, always claiming they were with the Chief. Lejeune ended up taking full credit for the results at a congress in August 1958 in Montreal, Canada. Gautier was only told about the 1959 paper shortly before it was submitted [24-26]. Lejeune received all the recognition for this discovery with Gautier given the backseat. However, eventually the truth prevailed, with Gautier receiving official recognition from Inserm's Ethics Committees – a French national institute for health and medical research. According to the committee summary: (1) Lejeune's part in the discovery was unlikely preponderant, (2) while his involvement had garnered international promotion, this is distinct from contributing to the discovery itself, and without the discovery the promotion could not have existed, (3) the technical approach is a necessary condition for discovery, and thus it was regrettable that Gautier (and Turpin's) names were not systematically associated with the discovery in communication and honors, and (4) Gautier's case is important in illustrating giving recognition where it is due [27]. This was a fair, just and overdue recognition of Gautier's work. In addition to her discovery, Gautier has other notable career achievements, such as being the founder and director of the Department of Anatomopathology of Childhood Liver Diseases at the Kremlin-Bicêtre Hospital, as well as being the Senior Scientist in 1967, then Director of Research at Inserm [28].

However, despite this recognition, Lejeune's pseudo-legacy stalked and suppressed Gautier's prominence. This was most evident when the French Federation of Human Genetics were going to honor Gautier at the conference *Assises de Génétique Humaine et Médicale* in Bordeaux on January 31st 2014. She was expected to give a speech. The morning of the conference, Gautier discovered the event was cancelled, as two bailiffs at the order of Jérôme Lejeune Foundation wanted to record the speech in the event Gautier would "tarnish" the good name of Lejeune. She ended up receiving her award unceremoniously a day later, with the conference organizers sending a notice of their regret at the cancellation and condemned the use of legal action [29]. What is particularly disturbing about Lejeune's foundation is the re-posting of a September 2018 report about Lejeune approximately two months after the passing of Gautier (she passed away April 2022, they posted the report June 2022). The introduction of their report still attested to the rightful claim being given to Gautier, referring condescendingly to the ruling of Inserm's Ethics Committee September 14th, 2014 deliberation as being "their opinion" and trying to defame Gautier's statements in the matter as having "many inconsistencies" [30]. However, given we are honoring the legacy of Gautier and those who found themselves in similar situations, *their* attempt at tarnishing Gautier has not succeeded.

On a more favorable note, this quote from Gautier's autobiographical paper tickled us: "I started on a PCB (first medical degree): easy enough."[24] Only a genius would find the study of medicine "easy enough".

# Kathleen Lonsdale

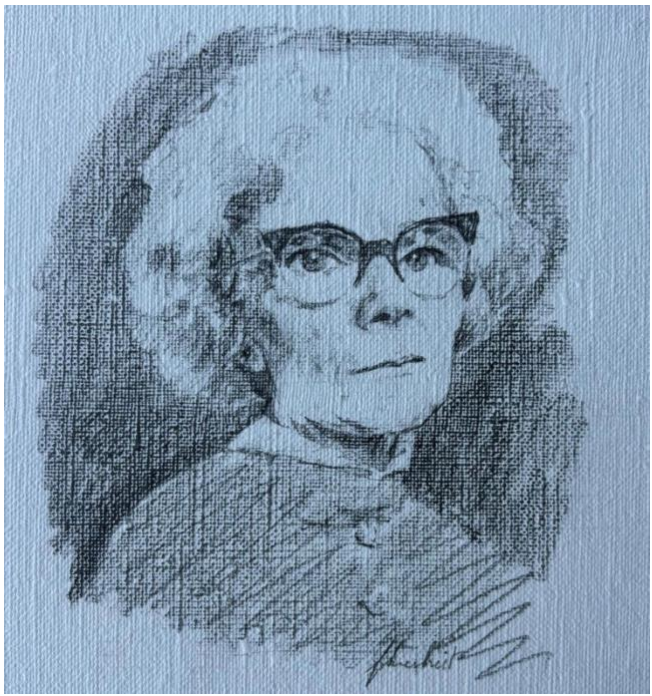

**Figure 11. Kathleen Lonsdale.** The gifted Lonsdale demonstrated the structure of benzene. However, despite her successes, she still faced gender-related biases in her career. These events fueled Lonsdale's advocacy for women in science. [*Illustrated by Farah Ajili, contributing author, 2024*]

*"Observation is not enough, and it seems to me that in science, as in the arts, there is very little worth having that does not require the exercise of intuition as well as of intelligence, the use of imagination as well as information."* – Kathleen Lonsdale

Dame Kathleen Lonsdale (née Yardley; **Figure 11**) was born in Newbridge, Southern Ireland on 28[th] January 1903. The youngest of ten children, Lonsdale's upbringing was filled with struggles, coming from a poor and broken-down family with a negligent, aggressive, and alcoholic father. Her mother eventually took her and her siblings to England in 1908, settling in Essex. Lonsdale initially was enrolled in the County High School for girls. However, she was later enrolled at the County High School for boys to undertake subjects in physics, chemistry, and higher mathematics (as these were not available in the all-girls equivalent school). Notably, she was the only female to have undergone this transition at the time.

Throughout her school life, Lonsdale's hard work and exceptional memory saw her earning several County scholarships. The Essex Education Authority offered to increase her scholarship if she continued to stay at the County High School, with the intention of getting Lonsdale admitted to Cambridge University. However, Lonsdale was anxious to leave, and ended up being enrolled at the Bedford College for Women in London – at the age of 16 years. At the college, Lonsdale initially studied mathematics but changed to physics at the end of her first year. While she was discouraged from making this change, with naysayers stating that Lonsdale had little chance of distinguishing herself in physics (little did they know), Lonsdale persevered, as she was worried that a career in mathematics would only lead her to a teaching position. Lonsdale ended up receiving first-class honors in physics, being one of eight students to achieve this feat. This achievement caught the attention of one of her examiners - crystallographer W.H. Bragg. Bragg offered her a place in his research team at University College London (UCL) [31, 32].

Lonsdale married Thomas Jackson Lonsdale in 1927 and moved to Leeds because of her husband's career. Her time in Leeds proved to be the fruitful. Richard Whiddington had worked on X-rays before W.H. Bragg, and thus welcomed Lonsdale to the Department of Physics. A part-time demonstratorship was arranged for Lonsdale to supplement her scholarship (the Amy Lady Tate scholarship by Bedford College). She was thus able to buy a new ionization spectrometer and electroscope with a grant of £150. The physics department was adjacent to the chemistry department lead by C.K. Ingold – who was researching aromatic substitution prepared by his students and offered Lonsdale beautiful crystals of hexamethylbenzene – her starting material. It was at this point that Lonsdale demonstrated the structure of the benzene – one of the building blocks of life - confirming its flat hexagonal structure using X-ray diffraction methods [33, 34].

Despite her groundbreaking work, and more favorable circumstances later in life as compared to other female colleagues, Lonsdale still faced her fair-share of gender-based barriers. Her initial application to the Royal Society was turned down, a decision influenced more by her gender than her scientific achievements. However, she continued her research undeterred. In 1945, she became one of the first two women to be elected a Fellow of the Royal Society, alongside biochemist Marjory Stephenson [35]. Lonsdale was also a committed pacifist, a stance that led to her imprisonment (spent at London's Holloway Prison for Women) during World War II for refusing to register for civil defense duties, such as fire-watching against incendiary bombs [36]. Her one-month-long imprisonment brought attention to the issue of conscientious objection.

Apart from her research, Lonsdale was deeply committed to promoting women in science. She was an advocate for work-life balance, especially for women scientists, and worked tirelessly to break down the barriers they faced. Lonsdale's articles and speeches on women in science in the 1960s were impacted by her own personal experiences. While her colleagues, such as Franklin, struggled against a different set of prejudices and assumptions as single female scientists, Lonsdale would face her own set of assumptions and obstacles as a married female scientist, with insinuations that her husband and family should take priority over her laboratory [37]. In their family dynamic, husband Thomas realized early on that his wife was an academic powerhouse, not shying away from taking on more domestic responsibilities that increased as his wife gained national and international recognition, providing a supporting role and even joining her during her advocacy for peace and penal reform [32].

Her contributions to crystallography and her efforts to improve the status of women in science were finally recognized later in her career. She was appointed Dame Commander of the Order of the British Empire in 1956 and continued to work and inspire until her death on April 1, 1971 [31].  Lonsdale's legacy is a testament to her enduring contributions to science and her role as a trailblazer for women in scientific research, overcoming the challenges posed by the societal norms of her time.

# Marguerite Perey

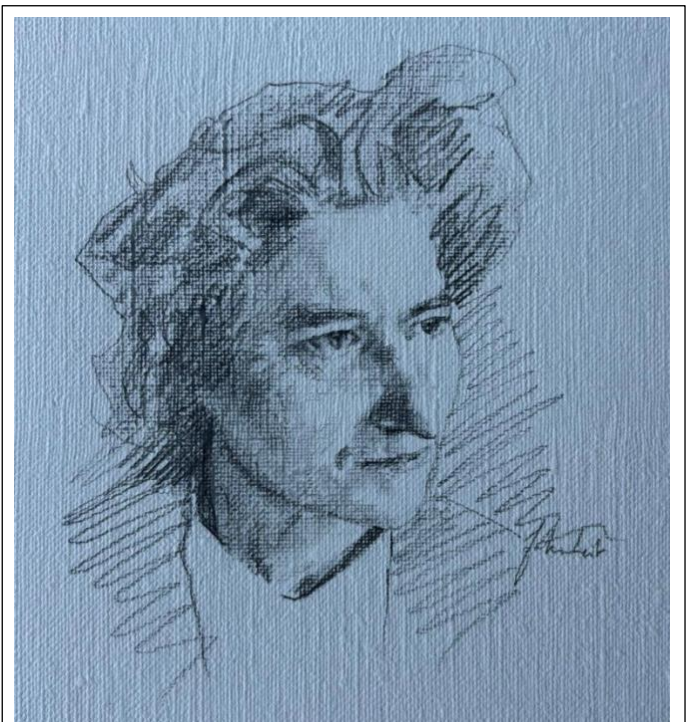

*"The most elusive element of all, however, appears to be francium, which is so rare that it is thought that our entire planet may contain, at any given moment, fewer than twenty francium atoms."* – Bill Bryson

**Figure 12. Marguerite Perey.** Perey discovered the element francium – named after her native France – in 1939 as a laboratory technician. [*Illustrated by Farah Ajili, contributing author, 2024*]

Marguerite Catherine Perey (**Figure 12**) was born 19th October 1909 in Villemomble, a suburb of Paris. The youngest of five children of a middle-class family, Perey hoped to make medicine her profession, but the socioeconomic state of her family prevented her from pursuing this career. She ended up studying chemistry at École Féminine d'Enseignement Technique, which was a school for laboratory technicians in Paris. It is here her fate collided with Marie Curie, who regularly hired the best chemistry laboratory technician from the graduating classes, where funding permitted. In 1929, Perey graduated top of her class, and was subsequently interviewed and hired by Curie, thus starting her career at the Radium Institute. Her primary tasks as Curie's laboratory assistant included getting the purest form of actinium [38].

Following the death of Curie, Perey reported to Curie's daughter, Iréne Joliot-Curie, and the new Director of the Institut du Radium – André-Louis Debierne. Unbeknownst to each other, both had asked Perey to determine the precise half-life on actinium-227. It was this work that inevitably led to Perey's discovery of francium (element 87) in 1939 – five years following the death of Curie. At the time of artificial radioactivity discovery, there were four very obvious absences from the periodic table, with little doubt that these spaces should be occupied. Francium was the last natural element to be discovered and was located between radon and radium [39-41]. Furthermore, it is considered rare, with only about 30g in the whole of the Earth's crust available [42].

An interesting disparity in historical readings suggests there were two differing original names for element 87. Some suggest the element was originally named "virginium", while others suggest "catium", with the latter being opposed by Joliot-Curie herself. Joliot-Curie's preference, it seemed, was the name thought up by Perey - francium - affectionately named after her native France [38, 41]. Furthermore, given that both Curie-Joliot and Debierne had given parallel directions to Perey, when the discovery of francium came about, both initially tried to lay direct claims to the work as Perey's primary supervisor. When the contention did not settle, it was instead agreed that Perey, as the discover, would be listed as the sole author of her 1939 publication [43].

Perey made this discovery as a modest 29-year old without a university degree, making a discovery that was missed by those with more experience and expertise [43]. In consideration of her success and talents, Debierne and Curie-Joliot encouraged Perey to obtain a formal qualification in science. She received her Doctorate from the University of Paris in 1946 as a result of her discovery of element 87. Following the completion of her degree, Perey worked as an independent researcher for "Centre National de la Recherche Scientifique" (CNRS), then eventually was appointed as the Chair of Nuclear Chemistry at the University of Strasbourg. At this point, Perey wanted to use her discovery of francium for early diagnosis of cancer [38].

Perey had the honor of being the first woman to be elected to the "Académie des Sciences" (i.e., the Physics section in 1962) – a privilege denied to Curie herself [44]. That said, this supposed honor bestowed upon Perey was incomplete. In the words of Adloff and Kauffman, "Perey was selected as a corresponding member….Such second-rank members have no 'academician seats' or other prerogatives of full members and do not bear the official title of *académicien*" [38]. Perey eventually ended up being the first woman full member in 1967 to the French Académie des Sciences [45]. This achievement garnered Perey considerable attention, as is illustrated in **Figure 13** [46].

> ▶ Mlle. MARGUERITE PEREY, director of the chemistry department, Institute of Nuclear Studies, Strasbourg, France, and known for her contributions to cancer research, has been elected as a corresponding member of the French Academy of Sciences. She is the first woman to be elected in the Academy's history of over 200 years. She was a laboratory assistant of Mme. Pierre Curie, whose admission to the academy was always refused. (*The Times*, March 13.)

**Figure 13. A snippet from the British Medical Journal 1962 [46].** The excerpt highlights Perey's achievement as the first woman to be elected to the academy, noting this honor was refused to her mentor Marie Curie.

What is most fascinating about this is Perey created more of a frenzy over her membership of the Academy than her discovery of francium. Her discovery remained largely unknown or ignored by the public. However, the media frenzy around her admission to the Academy was baffling to say the least. "First Woman Academician" read the headlines. There was also discussion about the dress code of a female academician. Newspaper columnist André George asked whether the presence of Perey would mean the easier admission and acceptance of other female academics. This did not turn out to be the case, as the next elected member appeared on 13 March 1978 in the form of physicist Yvonne Choquet-Bruyat, who was elected to the section of mechanical and computer science [47]. Perey's inauguration into the Academy ended up opening a media flood gate that was filled with debris: from being referred to as a martyr in science, having misleading information published about her path in academia (e.g., dates of her discovery and her thesis completion being completely falsified), to also having several rumors spread about her (e.g., she underwent 12 operations, she lost her ability to use her left hand because of her work with Curie in radioactivity, etc). The focus became less on her actual achievements, and more on a fictionalized version of her life (example can be found in **Figure 14**) [47]. Irrespective of the struggles, Curie and Perey were praised by feminist groups at the time as the symbols of emancipation for women in sciences.

In the midst of the plethora of rumors that had spread about Perey and her condition as related to radioactivity, eventually these rumors did manifest into Perey's real-life. Eventually and unfortunately, the side effect of Perey's work caught up with her. She shared the same ill-fate that was experienced by her predecessors within the Curie group, and developed cancer as a result of her work in radioactivity. She suffered 15 years from radiation-related diseases until her passing on 13th May 1975 [38]. Despite her achievements, Perey was nominated at least five times for a Nobel prize but was never its recipient [48].

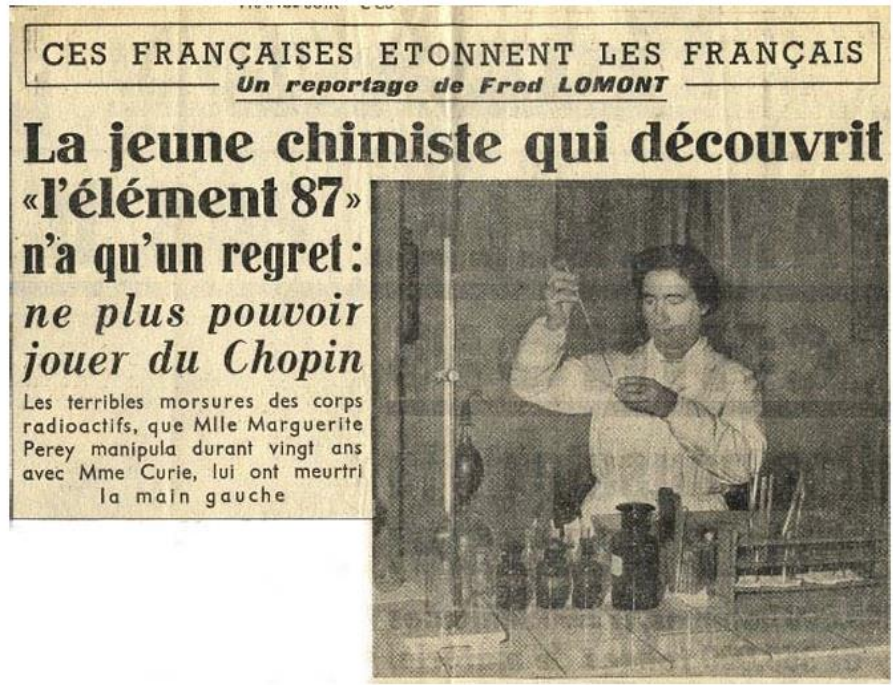

**Figure 14. Media Frenzy about Perey [47].** This piece reflects a fabricated report, claiming Perey's greatest regret was not being able to play Chopin again because of her exposure to radioactivity.

# Alice Ball

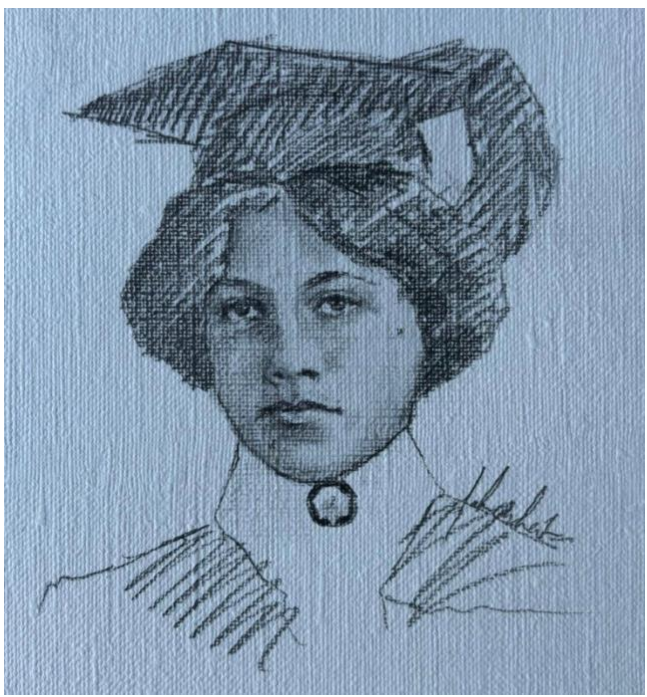

**Figure 15. Alice Augusta Ball.** The inventor of the *Ball Method*, Ball discovered a way to treat leprosy, in which she separated the active ingredients of chaulmoogra oil. This treatment halted the progression of leprosy while avoiding side effects which were previously an issue.

[*Illustrated by Farah Ajili, contributing author, 2024*]

*"Ball's discovery was very beneficial to alleviating pain that was sustained by patients…and for a black woman to be able to achieve what she did and make advances in that area during that time is remarkable unto itself."* – James P. Harnisch

Alice Augusta Ball (**Figure 15**) was born in Seattle in 1892 and graduated with degrees in Pharmacy and Chemistry at the University of Washington in 1911 and 1914, but most of her higher education was undertaken in Hawaii. She was the first African American woman to receive a Master's degree from the University of Hawaii and later became the first African American female Professor of Chemistry at the same university.

After finishing her graduate degree, she explored options for a Master's degree at various universities and was offered positions, including a place by the University of California. After completing her Master's degree at the University of Hawaii, she was approached by Harry Hollmann at the Kalihi Hospital and offered a position as an assistant in research into the treatment for leprosy, also known as Hansen's Disease. Until 1966, thousands of victims from around the South Pacific were interned in Hawaiian facilities.

In the past, the treatment for leprosy was based on using chaulmoogra oil, which was extracted from seeds of the tree *Hydnocarpus wightianus* [49]. The conventional applied treatment was attributed to Frederic Moaut in 1854 and had many problems, including difficulty of application and many side effects, such as vomiting or blistering. Alice Ball developed a new method that was much more effective and without many side effects. Her method separated the active ingredients from the chaulmoogra oil and produced a simple and convenient formulation suitable for injection without the problem of side effects [49, 50]. This treatment was known as the *Ball Method*. While not curing the disease the method could halt progression and was the best treatment until antibiotics were introduced [51]. The results raised her profile immensely but not without drama.

She was unable to publish her research before her death in 1916 at the age of 24 due to a laboratory incident involving chlorine gas contamination. To add further insult, Arthur Dean, a colleague and later dean of the college carried out further trials and published details of the work without any acknowledgements whatsoever, and the work was actually named the Dean Method [49-52]. Her name is not mentioned in any of Dean's published works on the chaulmoogra extract, while the name "the Dean method" was appended to the technique. This miscarriage of justice was eventually addressed by a fellow academics in subsequent publications [49, 50]. In 1922 Hollman released a paper praising Ball and naming the method after her, with other researchers consolidating her claim to fame [52].

Mushtaq and Wermager described her innovation involving the extraction of the water-soluble ethyl esters of chaulmoogra oil as having a profound impact on millions of people affected by Hansen's disease, who were socially shunned and confined to leprosy colonies [50]. Her experience mirrored that of other female scientists who were relegated to the shadows despite their great achievements, ironically at the hand of predatory male collaborators.

# Cecilia Payne

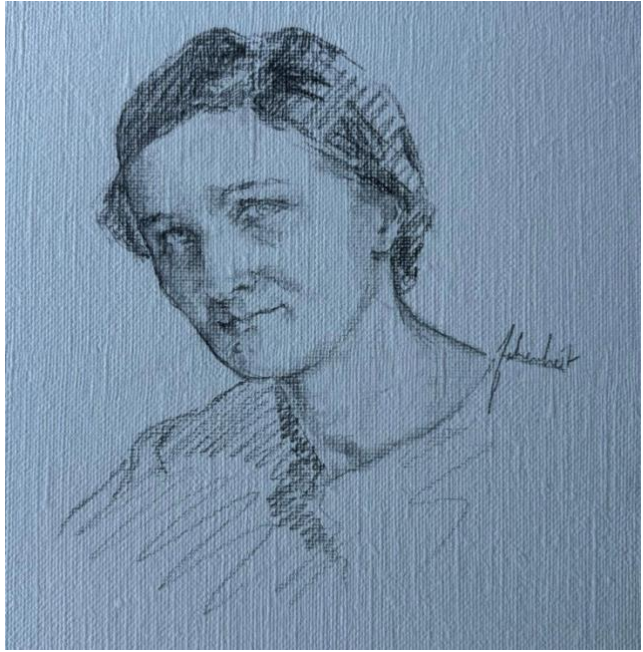

**Figure 16. Cecilia Payne-Gaposchkin.** Payne-Gaposchkin studied the element composition in stars, finding that helium and hydrogen are the most abundant elements in the universe. [*Illustrated by Farah Ajili, contributing author, 2024*]

*"Your reward will be the widening of the horizon as you climb. And if you achieve that reward you will ask no other."*

– Cecilia Payne-Gaposchkin.

Cecilia Payne-Gaposchkin (**Figure 16**) was born in Wendover, England in 1900. As a student, she was inspired by a lecture given by Arthur Eddington describing his expedition to the west coast of Africa to observe a solar eclipse to test the theory of relativity proposed by Albert Einstein [53]. This experience prompted her great interest in Astronomy. She finished her studies after many years - but was not awarded a degree by Cambridge University until 1948 because she was female and therefore not eligible. Due to limited career prospects, she left England for the United States in 1923 and subsequently gained a PhD, which was a first for a woman in Astronomy from Radcliffe College of Harvard University [54]. She became a US citizen in 1931.

She studied high luminosity stars and the Milky Way Galaxy. At that time, the prevailing belief was that the Sun and the Earth were similar in their percentage composition of elements. Her PhD studies resulted in a number of findings, including the conclusion that, in fact, the sun was composed mainly of hydrogen and helium. Moreover, this was also the case with other stars and therefore it followed that hydrogen was the most abundant element in the universe.

This conclusion was heavily criticized in a manner very similar to the treatment received by Galileo after his revelation that the Earth orbited the Sun, not the reverse case, which resulted in his persecution by the members of the church hierarchy. In response, she initially described the result as spurious until a prominent early critic finally agreed with her after validation by a different method but he was subsequently often credited with this discovery [55]. Despite this early set-back, her work was highly praised, but her career was still plagued by barriers due to the low status of women in academia and junior and low-paid positions [56].

Her later career and research were pioneering in its scope in field of stellar evolution and she was noteworthy for her teaching and publication [54-57]. She mapped thousands of stars in the Milky Way Galaxy and her studies helped to map the course of theory in stellar evolution. She was the PhD supervisor of the famous astronomer Frank Drake, known for the Drake Equation. She was scientifically productive over a long career at Harvard as her affiliation.

At the start of residence, women were not allowed professorships at Harvard University. Because of this, she was relegated to low-paid research positions for many years. Harlow Shapley, the Director of the Harvard College Observatory, who hired her, became an advocate for her after witnessing her undoubted talents. In 1956, she was eventually appointed as the first woman full professor at the Faculty of Arts and Sciences at Harvard University [54], with strong support from Don Menzel, the Director of the Harvard College Observatory. She later became the first woman to achieve the position of Head of a department at Harvard. She also encouraged other universities to admit more women in their science courses.

After a long and productive career at Harvard University, in 1966 she retired from teaching and was appointed Professor Emerita. She continued her research at the Smithsonian Astrophysical Observatory, as well as editing the journals and books until 1976. Payne died in 1979 at age 79 after a long and productive career and an advocate for more women in science programs in American universities. She shared many career similarities with other very talented women who struggled against an old-fashioned male-dominated culture in scientific research. Her example provided inspiration for other scientifically gifted women and here advocacy helped to gradually open-up career paths for young women in conservative education systems. She was a mentor and supervisor for aspiring young female scientists.

# Emmy Noether

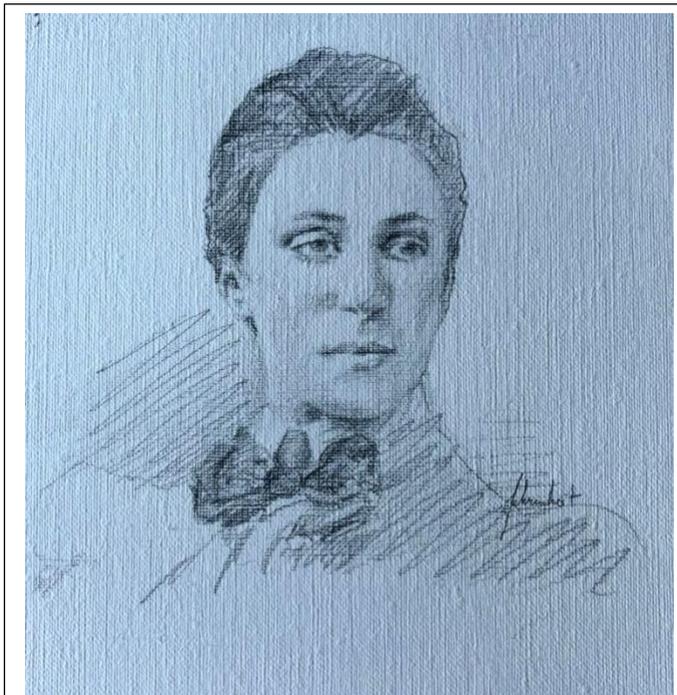

**Figure 17. Emmy Noether.** Described by Albert Einstein as "the most significant creative mathematical genius", Noether's contributions in mathematical physics, particularly Noether's first and second theorems, shows symmetry in nature leads to a conserved quantity. [*Illustrated by Farah Ajili, contributing author, 2024*]

*"My methods are really methods of working and thinking; this is why they have crept in everywhere anonymously."* –

Emmy Noether

Emmy Noether (**Figure 17**) was a mathematical genius who produced important contributions to abstract algebra and mathematical physics. This was despite working under pseudonyms because of discrimination based on being female while working in mathematics, which was regarded as a masculine discipline in the early 20$^{th}$ Century. She was born in Erlangen in Bavaria in the German Empire in 1882.

She gained her doctorate in 1907 from the University of Erlangen and worked at the Mathematical Institute of Erlangen without pay for seven years. This travesty occurred because females were prohibited from academic positions at that time. In 1915, she was invited by famous mathematicians David Hilbert and Felix Klein to the University of Göttingen, a world-renowned center for mathematics research, but due to objections by academics, was forced to work for four years as a lecturer under the name of Hilbert.

After her death, Albert Einstein wrote a letter to the New York Times in May 1935 describing her as "the most significant creative mathematical genius thus far produced since the higher education of women began…. in the realm of algebra, in which the most gifted mathematicians have been busy for centuries".  Many tributes were also provided by other eminent scientists, such as Norbert Wiener and Hermann Weyl.

The research produced by Noether was both innovative and prolific, making contributions to theoretical developments in general relativity, algebraic invariants, algebraic topology, number rings and fields [58]. Her work on differential invariants in the calculus of variations was described as one of the most important theorems ever proved in guiding the development of modern mathematical physics [59]. Her contributions include Noether's first and second theorems in mathematical physics. After 1919, she became a leading member of the Göttingen mathematics department until 1933.

She demonstrated an enormous capacity for abstract thought, which enabled her to approach various problems in physics and mathematics in new ways. Until 1919, she was mainly involved in research in differential and algebraic invariants. Subsequently she became familiar with the work of David Hilbert and was greatly inspired by his work. Following her move to Göttingen in 1915, she produced her famous work in physics, the two Noether's theorems. In the period 1920 to 1926, her research focus was the theory of mathematical rings [60]. From 1927 to 1935, her focus was on noncommutative algebra, linear transformations, and commutative number fields [61].

She was forced to flee as a refugee from Nazism in 1933. She was invited to Bryn Mawr College in the United States and received enthusiastic support from staff and admirers of her achievements. This was finally a pleasant time in the social context with appreciation from staff and female academics. In 1934, Noether was an invited lecturer at the Institute for Advanced Study in Princeton. She felt even at Princeton in the new world that she was not welcome being a female academic [62]. Nevertheless, she became a celebrity at Bryn Mawr College, supervised doctoral students in the US and Germany and was a leading figure and role model.

She died in 1935 aged 53 at Bryn Mawr in Pennsylvania and her ashes are interred at Bryn Mawr College.

With respect to recognition, she was ranked by peers as one of the greatest mathematicians of the twentieth century and even the greatest female mathematician in recorded history [62]. She provides a role model and inspiration to all women in the STEAM disciplines and in particular, mathematics and physics. It was particularly impressive that she achieved so much while suffering from unjust discrimination throughout her career. Despite all her achievements she remains largely unknown to the general public.

# Flora Tristan

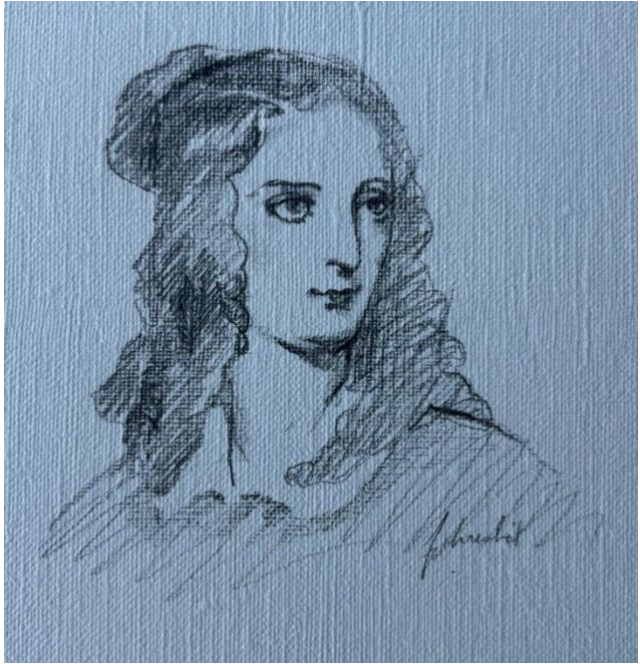

**Figure 18. Flora Tristan.** French-Peruvian Tristan had a tumultuous life which culminated in her criticizing the social and economic injustices of her time. Her self-published book *The Workers' Union* had made its way into the world five years prior to Marx and Engels publication *The Communist Manifesto*, yet her legacy and presence remain largely unknown. [*Illustrated by Farah Ajili, contributing author, 2024*]

*"The Woman is the Proletarian of the Proletariat"* – Flora Tristan

Before Marx had yet set pen to paper for the first time, Flora Tristan (**Figure 18**), née Flore Célestine Thérèse Henriette Tristán y Moscoso had begun drawing the connections between capitalism and the oppression of the Proletariat. Tristan published her central work, *The Workers' Union* five years before Marx and Engels published *The Communist Manifesto.* While Engels and Marx paid tribute to other 'Utopian Socialists' in *The Communist Manifesto,* Tristan's name never made it onto those pages. The French-Peruvian socialist activist had found a handful of truths in her travels through France, Peru, and England, with the main conclusion being that the political and economic systems du jour seemed to be functioning as a means of oppression for the working class and that working conditions would never be improved by the bourgeoisie and the Proletariat would have to take their liberation into their own hands. Few truths can in all earnest be considered universally acknowledged, but as Tristan travelled from one part of the world to the other, she found herself noting the strife of the working class and the subjugation of women in all social classes. It did not seem to matter how disempowered a man you would find in any social group, he would always be more powerful than at least one person (his wife). As an avid watcher and supporter of the French Revolution, Tristan found it difficult not to notice how "the rights" they were fighting to pass through parliament were only "The Declaration of the Rights of *Man* and of the Citizen*.*" She could not help but note the fact that their collectively hard-won *liberté* only had eyes for the *fraternité*.

Flora Tristan was born in Paris in 1803 to her Peruvian father and French mother. When Tristan was 5, her father passed away unexpectedly, which was the catalyst for a life full of turmoil, strife, and tireless ambition. Upon her father's demise, it was discovered that her parents' marriage was not considered legitimate by either Peruvian or French standards, essentially granting Tristan and her sister the titles of 'bastards.' Tristan's father had come from one of the wealthiest Peruvian families, which had afforded them a very nice lifestyle. However, with an illegitimate marriage, neither their widow mother nor the children were entitled to any inheritance. After this, they were forced to move out of their Paris mansion and into the slums surrounding Paris.

When she turned 17, Tristan married her husband out of a need for stability. As it turned out, he was both drunk and abusive. By the time Tristan was 22, she already had two children and was expecting a third. At this time, her husband tried to send her into sex work against her will. It was this event that ultimately spurred Tristan on to take her children and leave. However, this was a time wherein divorce was neither legal nor a real option, so Tristan suddenly found herself doubly the pariah, both as a 'bastard child' and as a 'woman of divorce.' It was at this particular intersection Tristan, for the first time, found herself in the situation of not having a single discernible right within any social or legal system.

Knowing that she had to work and knowing that being a single mother of three would not get her any jobs, she sent her children to a boarding home and started working as an escort to women who were otherwise traveling by themselves. This served two purposes: 1) she managed to earn a decent wage while getting to travel, and 2) it took her out of Paris, where her husband was furiously searching for her. Her husband did locate and assault her multiple times up until 1838 when he seriously wounded Tristan after shooting her twice. This finally earned him 20 years of compulsory labor. Some scholars speculate that her husband only got such a severe punishment because by then, Tristan had already published her first booklet, which had gained quite a bit of attention and earned her a reputation [63].

In 1833, Tristan travelled to Peru to appeal to her father's family for her inheritance. They agreed to an annual stipend, which was equivalent to 6% of what she would have inherited as her father's legitimate daughter. Upon her return, Tristan wrote *The Necessity of a Pleasant Reception of Foreign Women,* which was her first published booklet. Herein, she argued in favor of practical traveling accommodations and facilities for traveling women. What made Tristan's work unique was that she encouraged women to band together and collaborate in their fight for equal rights rather than attempt to appeal for change with the existing institutions of power. This type of thinking is what we in modern feminist theory consider radical feminist ideology [64]. However, it is worth noting that the word 'feminism' was not yet invented at Tristian's time, and so, therefore, she did not identify as such. One of Tristan's earliest realizations in her work was that the liberation of *any* oppressed group would have to come from their own hands, as no oppressor had any interest in the liberation of the group of people whose oppression they benefitted from. This also included women. In fact, Tristan famously noted that "The Woman is the Proletarian of the Proletariat," herein tying together the issue of social class with gender [65]. What Tristan argues is that within every social class and group, women always come out as the lowest-ranking members. You could be the poorest, least powerful man in the world, but you would still rank higher than your wife based on no other merit than that she was a woman. So, while Tristan did not identify as what we today call 'feminist,' she did find that there was an inextricable link between the liberation of the working class and the liberation of women. Since women made up a substantial amount of the working class, it seemed only logical to Tristan that they should be included in its liberation. These beliefs did not stop at women and laborers, and in 1838, Tristan published *The Peregrinations of a Pariah,* wherein she strongly criticized the Peruvian oligarchy and the Peruvian government's laws, commitment to maintaining social and economic injustices, as well as its involvement with slavery. When her Peruvian family learned of the book, they were all horrified by its content and cut her off financially.

Tristan, being as difficult to deter as ever, decided on a new venture and travelled to England, where she interviewed prisoners, sex workers, and homeless people and visited factories, slums, and prisons, all in an effort to assess the socio-economic state in England. After her thorough exploration of English society, she published the work *Walks in London* in 1840, wherein she concluded that exploitation was the main cause of unhappiness amongst working men and women. In 1843, Tristan finished her last and most well-known work, *The Workers' Union*. However, no publisher would dare to go near it, so Tristan decided to self-publish once more as she had with her first work. She needed to raise the money on her own, and so she decided to find subscribers for the book. In Paris, she contacted approximately 200 people and many more by letter. She managed to find enough subscribers for the first and second printing of the book, selling mainly to the bourgeoisie. She printed 4000 copies of the first edition and ten thousand of the second edition. By the third printing in 1944, she printed ten thousand more, and unlike the first two editions, these were almost exclusively bought by working-class individuals.

Having received over 200 letters from workers about her writings and their experiences, she decided to travel to where the workers were. She travelled throughout France, where she met with workers and read to them from her book. This was not without difficulty, as she frequently experienced the police attempting to apprehend her and stopping her from attending the gatherings. The very picture of perseverance, Tristan never once let it stop her, and according to historical contemporaries, Tristan one time lost her temper when the police rented the room next to hers in a hotel in order to spy on her. This infuriated her to the point where she confronted the officers and demanded that they either come back with a warrant for her arrest or stay away altogether [66]. Tristan, having never slowed down a day in her life since her father died at age 5, suffered from exhaustion and finally passed away from what is believed to have been a heart attack in Bordeaux in 1844 aged only 41 [67].

# Jocelyn Bell Burnell

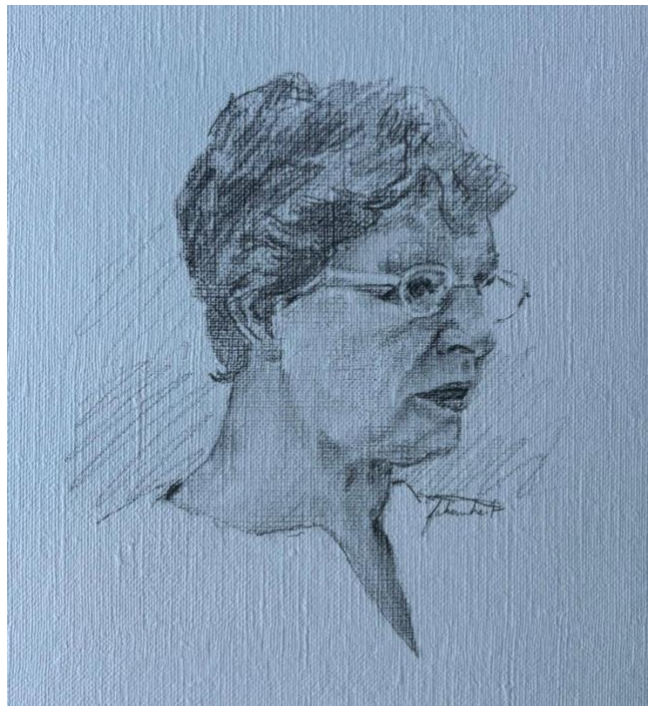

**Figure 19. Jocelyn Bell Burnell.** During her PhD studies, Bell Burnell discovered pulsars – pulse of radio waves emitted by a dying star. Her discovery was initially dismissed by her supervisor, Antony Hewish. Ironically, when her discovery was validated, Hewish took claim and credit for her work, eventually being awarded the Nobel prize while Bell Burnell was pushed to the curb on her own discovery. [*Illustrated by Farah Ajili, contributing author, 2024*]

*"That settles it! It's man-made"* - Antony Hewish, Jocelyn Bell Burnell's PhD supervisor

This was the discouraging feedback that Jocelyn Bell Burnell (**Figure 19**) received from her PhD supervisor Antony Hewish, back in 1967, after identifying for the first time a pulsar signal in a radio telescope built by her. Seven years later, the Nobel Prize in Physics was awarded for the first time in the astronomy field for "pioneering research into radio astrophysics and the discovery of pulsars". Antony Hewish and Martin Ryle were awarded; Jocelyn Bell Burnell was not recognized for her own discovery [68].

Jocelyn Bell Burnell was born in Northern Ireland in 1943 and from a young age she was already certain about two things: her love for astrophysics and the need to fight for it.

*"Women were not expected to be anything other than sex objects, wives, mothers, housewives... You were not expected to have any brain, you were not expected to have any career. Getting married was the goal"* - Jocelyn Bell Burnell, Journeys of Discovery interview by University of Cambridge, 2020 [68].

Her interest in astronomy started at an early age when she was about 10-11 years old. Her father was one of the architects of the local planetarium and would bring her home astronomy books from the library which triggered Bell's interest to become an astronomer. Motivated by her passion, she overcame her school's ban on girls studying science and discovered also her interest in physics [69, 70].

At 18 years old, unsure whether to study astronomy or physics, she decided to enroll at the University of Glasgow since it had both degrees. She ended up choosing physics, as she considered it to be a more versatile discipline. However, the path ahead of her was not easy. Universities and in particular some science domains such as physics were male-dominated and a hostile environment for female students. As an undergraduate at Glasgow, she was the only woman in a class of 50 students [70].

*"When a woman entered the lecture theatre, all the guys would whistle, stamp, beat on the desks and catcall. This happened at every lecture. If you blushed, they only made more noise, so I learned not to blush"* - Jocelyn Bell Burnell, Avenue interview by University of Glasgow, 2023 [70].

After her stay in Glasgow, she went to Cambridge to pursue a PhD in astronomy. For two years, together with five other colleagues, Bell built a massive radio telescope (equivalent to the size of 75 tennis courts), specially designed to monitor scintillation from quasars that were discovered just some years before [68].

This massive radio telescope became her responsibility. Her tasks involved distinguishing real galaxy signals from human-generated interference that is usually picked up by radio telescopes. On top of that, she needed to do all her analyses by hand and a complete scan of the sky would take four days and require 120 meters of paper. She analyzed kilometers of data [69].

While analyzing her data, Bell noticed something unusual, that she initially noted as "bits of scruff". She found a regular signal (a series of sharp pulses every 1.3 seconds) consistently coming from the same patch of sky, too fast and too regular to be made by a quasar. She knew this signal was distinctly unusual, so she reported it to her supervisor Antony Hewish who initially devalued her discovery, stating she was just measuring artificial radio interference [68].

*"I was quite convinced I was not clever enough to be in Cambridge. They would throw me out at some stage, but my policy was to work as hard as I could so that WHEN they threw me out, I'd know I'd done my best."* - Jocelyn Bell Burnell, Journeys of Discovery interview by University of Cambridge, 2020 [68].

Despite suffering major impostor syndrome, Bell Burnell persevered. She didn't know what these pulses meant but she was convinced they were not an artifact, as suggested by her supervisor. She continued to investigate them.

Doubtful of her discovery, Hewish pointed out the sporadic nature of the signal (it was only one part in ten millions of all the charts she had) and suggested her to try to capture it in another telescope, to rule out the possibility that she had wired the radio telescope wrong [68]. After seeing the pulses with his own eyes, Hewish decided to announce the findings by giving a seminar at Cambridge and by publishing the results as first author in *Nature* [71], diminishing the main role of Jocelyn Bell Burnell on the discovery and instead making himself the main actor. Further studies by groups of astronomers around the world identified also these kinds of signals [68].

The mysterious source of these signals rapidly gained interest within the scientific community but also in the press. The scientific community settled on the theory proposed by astronomers Baade and Zwick in the 1930s. This theory predicted that when a massive star died it would collapse into a super-dense shrunk neutron star that would spin very rapidly. According to this theory, the signal observed by Bell Burnell must be the pulse of radio waves emitted by this dying star which is why it was later called pulsar ("pulsating" + "star").

The press response was highly misogynist. Hewish was the main respondent to all the scientific questions while Bell Burnell had only personal questions directed to her, namely about her body and appearance (an experience still shared by women in the media today).

*"After the discovery, there were a lot of television interviews. There'd be my thesis advisor and myself, and they'd ask him about the astrophysical significance of the discovery, which he duly gave them, and then they'd turn to me for what they'd call the 'human interest,' like how tall was I, what were my bust, waist, and hip measurements, how many boyfriends did I have, that kind of thing. I wasn't a scientist; I was some kind of sex object."* - Jocelyn Bell Burnell, Interview by Ailís, Young Scientists Journal [72].

*"It was bizarre and it was offensive. And the newspaper photographers would ask me to undo some of the top buttons of my blouse. So I found that extremely difficult. But I felt I wasn't in a position to kick up a fuss. I was still only a student. I had to write a thesis, graduate and get references."* - Jocelyn Bell Burnell, CBC radio interview, 2018 [73].

*"I couldn't make too much of a fuss because I'd already invested almost three years in this project, I had six months to go and I wanted my PhD, please, by this stage. So I didn't feel I could rock the boat too much."* - Jocelyn Bell Burnell, Avenue interview by University of Glasgow, 2023 [70].

The discovery of pulsars was so important for the astronomical field that it was widely recognized in 1974 by the Nobel Prize Committee. Jocelyn Bell Burnell who was the researcher that constructed the radio telescope, discovered this sporadic signal in the huge amount of data she had to analyzed and argued and showed the signal was real and not an artifact, was at the end of that tiring journey not recognized [74].

After her PhD, Bell Burnell taught at multiple universities in the United Kingdom, including at the Open University. She studied the sky in almost every region of the electromagnetic spectrum. Jocelyn Bell Burnell took on leadership roles such as President of the Royal Astronomical Society, project manager of the James Clerk Maxwell Telescope in Hawaii and visiting professor at Princeton and Oxford. She was also the first female President of both the Institute of Physics and the Royal Society of Edinburgh. Bell Burnell became a member of the Royal Society in 2003 and in 2007, aged 64, she was made Dame Commander of the Order of the British Empire by the Queen. In 2008, se was elected to a two-year term as president of the Institute of Physics. In February 2013 she was assessed as one of the 100 most powerful women in the United Kingdom by Woman's Hour on BBC Radio 4. One year after, she was made President of the Royal Society of Edinburgh, the first woman to take that office [75].

In 2018, aged 75, more than 50 years after her discovery, she won the Special Breakthrough Prize in Fundamental Physics and she decided to donate the entire prize ($3,000,000) to create a scholarship fund for helping underrepresented students to study physics at the PhD level and therefore creating a space for diversity in physics to thrive [75].

*"Diversity adds to the creativity of a team, it brings an extra openness, and scientific breakthroughs are about taking data and when you come across something new, examining it open-mindedly. People from non-traditional backgrounds will not necessarily make the traditional assumptions, and that's how you get breakthroughs. That's what I did: I saw the data and realized it did not fit and needed attention – it was an anomaly that did not fit, and so was I!"* - Jocelyn Bell Burnell, IOP – The woman behind the fund [76].

In 2020, Jocelyn was one of several women to feature on a new £50 banknote in Northern Ireland and one year later she received the Copley Medal, the most prestigious scientific award in the United Kingdom, given annually by the Royal Society of London "for outstanding achievements in research in any branch of science [75].

Jocelyn's work was introduced to a new audience when a graphic of the pulsar she discovered became the artwork of Joy Division's debut album Unknown Pleasures in 1979. Peter Saville, Joy Division's album designer got inspired by *"seeing a stacked plot of radio signals from a pulsar in The Cambridge Encyclopedia of Astronomy"* [68]. The illustration became a musical icon.

It is insane the amount of difficulties Bell Burnell had to face just to pursue her scientific passions. Jocelyn Bell Burnell is an example of strength, perseverance and empathy. She is both a great scientist and a great human being.

# Marianne Weber

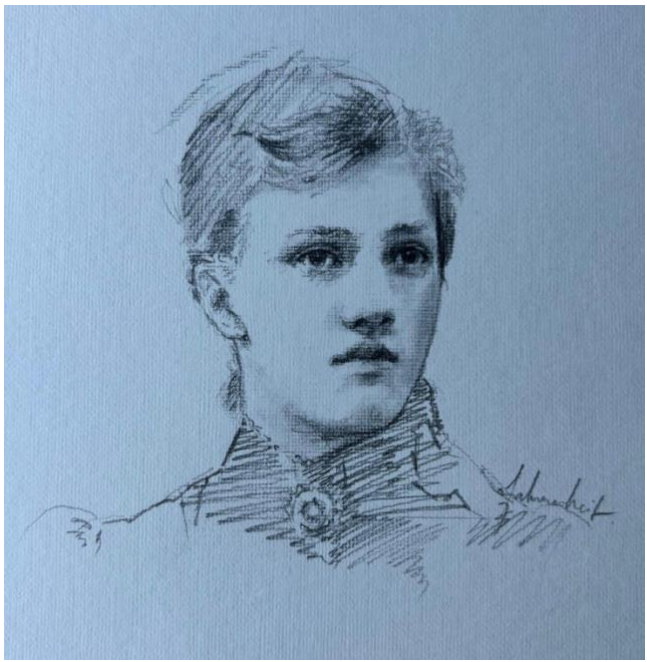

**Figure 20. Marianne Weber.** Despite her openness, courage, and conviction and her advocacy for equal rights for women in every aspect of life, writing the story of Marianne Weber proved to be most difficult in this series, as every piece of information that exists about her is connected to one of the men in her life, namely her husband Max Weber. [*Illustrated by Farah Ajili, contributing author, 2024*]

*"Marianne Weber is today recognized as the wife of a famous man and as a feminist of sorts"*

- Guenther Roth [77]

*"I don't know if it is permitted, but I feel great impatience about all that men state about us, including Georg . . . I wish that we would become again human beings, female human beings instead of exaggerated femininity, over-determined by an orientation to men, that we have practiced already about a thousand years and which I refuse to consider our female nature"*

– (Philosopher Getrude Simmel in a letter to Marianne Weber, discussing Simmel's husband's approach to 'the woman question' in his writings, which Weber contested in her own) [78]

*"Marianne Weber, widow of the famous sociologist"*

- [79]

To write this profile about Marianne Weber (née Schnitger; **Figure 20**) has proven a difficult task. Not only because what remains of Weber is limited to her publications and those of her letters and correspondence, but more so because by far and large every piece of information that can be found on Marianne Weber exists in relation to one of the men in her life. Largely, you will see her as a footnote to her, historically, more famous husband, Max Weber. Occasionally you will see her mentioned in connection with someone like Georg Simmel addressing her academic disagreements with him on 'the woman question'. All in all, it is rare that Marianne Weber is brought up solely with the purpose of discussing her works and her rather outstanding contributions to political and socio-legal academic thought. It is a peculiar case, that Marianne Weber has been so readily erased from history as a singular human being and only seem capable of existing in connection to a man, when she was so well-respected and outspoken at her time. As Ulrich [80] said "*Well-behaved women seldom make history*", it seems that being an outspoken, bold woman does not guarantee you any sort of capital in the hallowed halls of history either. Marianne Weber was never the sort to settle for less than what any man was offered, and so it seems largely tragicomic that her memory has been reduced to a footnote of her husband's name – the very sort of reduction of women's existence she fought against. Alas, this biography will not be free of neither Max nor Georg, as to explain the full extent of Marianne Weber's truly impressive academic skill and legacy, I must draw on them for context. However, this biography aims to center Marianne Weber and discuss her history, her person, and her academic work above all. While her thoughts might not be considered the most radical of feminist agendas today, for her time, her published works and political speeches were plenty radical, and represent the successful career of a woman who refused to settle for a quiet, second-tier life in the shadow of a man's.

Marianne Weber was born in 1870 into a family with moderate financial issues and some social standing. After her mother's death in 1873, her father sent her to be raised by her grandmother and aunt, where she stayed for the next fourteen years. At 16 she was sent to one of the more fashionable finishing schools where "she allowed herself to be 'finished'," learning to dance, to speak French and English, to appreciate music and art, while inwardly resolving "to become someone of note … [with] a 'burning ambition' to achieve" something that would make her "first in importance" [81]. She graduated in 1889, the same year her grandmother passed. After this, she moved in with her mother's sister, which is where she also met the man who would become her husband: Max Weber.

After marrying in 1893, Max and Marianne moved to Berlin, where Max worked at the University of Heidelberg. Marianne Weber herself started studying under the neo-Kantian philosopher Heinrich Rickert. This was despite the fact that she did not have any official exams, having attended a finishing school as her primary education. Still, she went on to be a part of the first generation of university women in Germany. Her first publication was "Fichte's Socialism and its Relation to Marxist Doctrine" in 1900.

In these years, Marianne Weber also got increasingly interested in feminism and politics. There was a handful of different strains of feminist ideology circulating at the time, and Weber belonged to one of the more dominant ones, namely, liberal feminism. Liberal feminism took up the issues of equal rights, such as demanding women also get access to education, that they receive equal wages for equal work, that they receive benefits for health and childcare, as well as them having full political rights. Moreover, they demanded changes to marriage laws, allowing women property rights, right to their children, access to divorce, and that the law should acknowledge marriage as an equal partnership rather than default to the woman as subordinate to the husband [81].

In 1907, Weber published her study "Marriage, Motherhood, and the Law", which was well received in feminist and political circles, and catapulted Weber into the position of one of the leading liberal feminist intellectuals. This success led to Weber being the spokesperson for the Federation of German Women's Organizations, one of the leading and most powerful liberal feminist organizations in Germany. She first served as spokesperson where she gave many lectures and speeches for the Federation, and in 1920 she went on to serve as President of the Federation. However, despite primarily looking through the lens of liberal feminism, she was also interested in and inspired by socialist feminism, cultural feminism, and the "erotic movement." For all of these she was inspired by parts of their arguments, but rejected or neglected others in her own work. For socialist feminism, she was inspired by the writings of the time, and acknowledged the issue of class in her own writings but did not draw connections between capitalism and the subjugation of women. Cultural feminism was the dominant feminist strain prior to liberal feminism in Germany, and acknowledged the unique cultural, spiritual, and ethical qualities inherent to women. While Weber acknowledged some cultural traits and struggles seemed inherent to women, like motherhood, she did not agree with the ideology's core claim about fundamental differences between the sexes. The "erotic movement" sought to reconnect men and women with their base human energies and free them from repression projected onto humanity by puritanical ideals. In its core, erotic feminism encouraged sexual experimentation and rejected monogamy. The movement believed women's liberation was through a critique of heterosexuality, exploration of the homoerotic, and dismantling monogamy. However, Weber believed in the 'intimacy' of marriage and that it should be a long-term project of a partnership, and thus rejected the idea of 'free love'. More importantly, Weber considered marriage as "as a complex and ongoing negotiation over power and intimacy" [81].

In her years with the Federation of German Women's Organizations they managed to achieve numerous significant advances for women in Germany, including seeing the first woman selected to parliament. Academically Weber was a significant contributor to the intellectual scene in Berlin. She was among the first generation of women allowed at German Universities where she studied under highly respected academics. Her work was considered thought provoking, well researched, and sound. She was known for arguing against male scholars' interpretation on the "the woman question", most noticeably her years long back-and-forth with Georg Simmel on how to address the inequity of labor in marriages. Moreover, Weber hosted a salon for intellectuals in Berlin which was well-known and well-respected.

After Max Weber's sudden and unexpected demise, Marianne Weber dedicated herself to finishing what was left of her husband's legacy and allegedly declared "Max Weber's desk is now my alter" [82]. In this time, she continued to be a woman of mettle, as she overturned the will she and Max had implemented just two years prior, and demanded full access and rights to his current work, including the unfinished manuscripts. Then she further shocked people by performing a speech in front of everyone at his memorial service, something that was not considered decent for women to do at the time [82]. Weber would then spend the next seven years of her life pulling together Max Weber's scattered notes and turning them into some of his most influential works. The manuscripts and projects that lay too far outside her expertise, such as his studies on ethnomusicology, she carefully entrusted in the hands of other experts in the field. Likewise, she offered her own work on Max's manuscripts up for critique and feedback from other renowned contemporaries in the field to ensure the quality of what was produced.

Throughout her political and scholarly careers, stretching roughly from 1900 to 1920, Marianne Weber published fourteen works, including her most famous piece "Wife and Mother in the Development of Law." After Max' death in 1920, she did not release any of her own work until 1926, where she released a biography of Max Weber. In the time she devoted to Max's intellectual legacy she finished *Wirtschaft und Gesellschaft* (The Theory of Social and Economic Organization) which is widely considered to be one of Max Weber's most famous and influential works, she edited four volumes of his collected works, and lastly assembled Max Weber's correspondence, in order to preserve his personal and scholarly memory. Following this, Weber went on to publish three more of her own pieces, namely *Die Frauen und die Lieben* (Women and Love) in 1935, *Erülltes Leben* (The Fulfilled Life) in 1942, and lastly *Lebenserinnerungen* (Memoirs) in 1948.

For the remainder of her life Marianne Weber remained in Berlin. After the extended period grieving Max Weber, Marianne Weber once more opened her salon and invited academic debate back in her house. Throughout World War II and the existence of Nazi Germany, Weber continued the salon and, alongside her compatriots, decided to not openly rebel against the regime, but keep their critiques between the lines. In 1945 Marianne Weber told Howard Becker in an interview "one after another of our acquaintances disappeared, never to be seen again", and further explained "None of us were of the stuff of which martyrs are made. Perhaps this is unfair, for there is no sense of being a martyr when there is nobody to witness the martyrdom and be affected thereby. […] We would have been quietly exterminated in the dark, so to speak, with no one to witness our agony" [79]. And if there was one thing about Marianne Weber, the woman who had once sworn to use her burning ambition to become someone of note, who spoke confidently of her thoughts in front of men and women alike, who passionately argued for women's right to be equal in the eyes of the law, who clawed her way tooth and nail from having no formal education to being a part of the first generation of women at German universities and furthermore contributed *multiple* of the most era defining and influential academic works, both under her own name and in her contributions to Max Weber's work, this woman was unwilling to disappear quietly into the dark shadow of any man.

# Chien-Shiung Wu

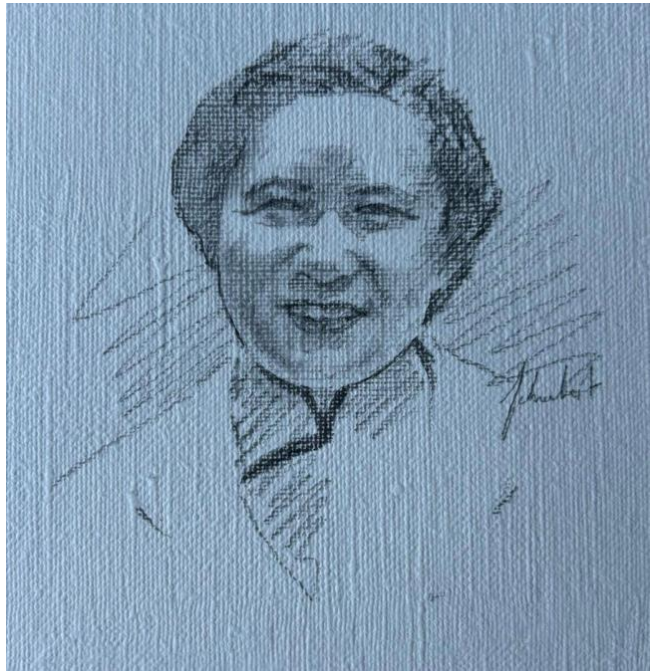

*"It is shameful that there are so few women in science."* – Chien-Shiun Wu

**Figure 21. Chien-Shiung Wu.** Despite Wu disproving the conservation of parity in weak interactions, she was excluded from the 1957 Nobel prize, which was instead awarded to Chen Ning Yang and Tsung-Dao Lee. [*Illustrated by Farah Ajili, contributing author, 2024*]

Parity symmetry states that physical laws remain the same when the spatial coordinates are reversed. Just as parity symmetry ensures that physical laws are the same regardless of spatial orientation, gender equality advocates for fairness and equal treatment for all individuals, irrespective of their gender. It is ironic that the scientist behind uncovering the law of parity symmetry, also recognized the impossibility of living up to its analogy. This is the story of The First Lady of Physics [83], who climbed up the mountains of physics but never got the recognition she deserved.

Chien-Shiung Wu (**Figure 21**), born in Liuhe, Jiangsu province, China, earned her bachelor's degree from National Central University in 1934. She initially considered attending the University of Michigan but discovered that women were not permitted to use the main entrance and had to enter through a side door. As a result, she opted to study at Berkeley instead where she obtained her Ph.D. in physics. Wu briefly taught at Smith College and Princeton University before joining Columbia University's Division of War Research in 1944 [84, 85].

She joined the Manhattan Project, a covert U.S. Army initiative focused on atomic bomb development. Within this project, she made substantial contributions, including solving a problem that had stumped physicist Enrico Fermi and devising a method to enrich uranium ore, crucial for producing bomb fuel [86]. In the mid-1950s, Wu was approached by Tsung-Dao Lee of Columbia and Chen Ning Yang of the Institute for Advanced Study in Princeton, N.J. They challenged the conventional belief in parity conservation, proposing that nature might distinguish between left and right in weak nuclear reactions. Given the challenge, Wu initiated a pioneering experiment to test this hypothesis. Utilizing a laboratory at the National Bureau of Standards in Washington, D.C., Wu conducted experiments with cobalt-60 in a strong electromagnetic field. By observing the behavior of nuclei as they decayed, she discovered a significant discrepancy: more particles ejected in the opposite direction of the nuclei's spin. This groundbreaking finding invalidated the conservation of parity law.

The discovery of parity violation holds profound implications, potentially shedding light on the universe's matter genesis. Subsequent experiments confirmed this phenomenon, cementing its significance in our understanding of fundamental physics [87, 88].

Despite her pivotal contribution to the discovery, Wu was excluded from the 1957 Nobel Prize awarded to Yang and Tsung-Dao Lee, reflecting the gender bias prevalent in that era. Wu addressed the issue during a symposium at MIT in October 1964:

"*I wonder whether the tiny atoms and nuclei, or the mathematical symbols, or the DNA molecules have any preference for either masculine or feminine treatment."* [75].

In her later life, Wu continuously opposed gender discrimination in science and criticized the repressive policies of the Chinese government. She passed away in New York City in 1997 at the age of 84, leaving behind a legacy of activism and scientific achievement [89].

# Elinor Claire Ostrom

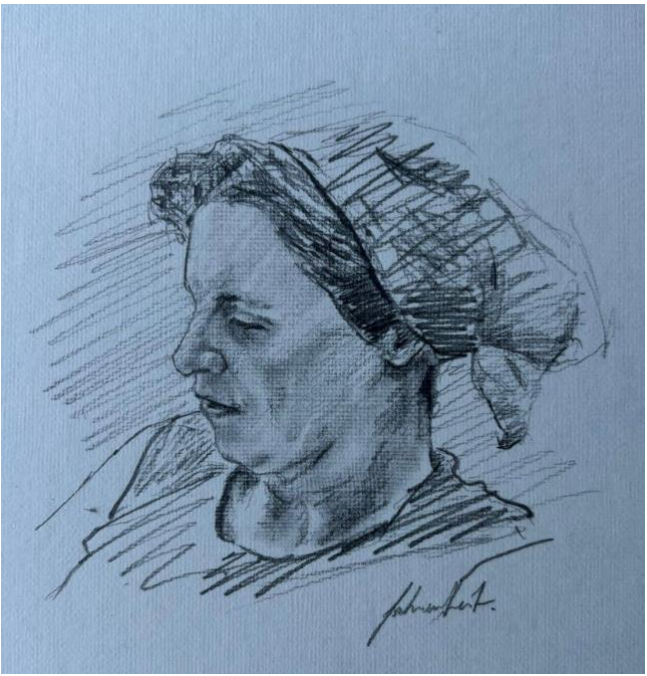

**Figure 22. Elinor Ostrom.** Despite her humble beginnings, Ostrom was the first women to win the Nobel prize in Economics in 2009. She challenged conventional wisdom, promoting community-led management of resources. [*Illustrated by Farah Ajili, contributing author, 2024*]

*"The power of theory is exactly proportional to the diversity of situations it can explain."* – Elinor Ostrom

Elinor Claire Ostrom (née Awan; **Figure 22**), born in 1933, became the first woman to be awarded the Nobel Prize in Economics in 2009. Raised in a modest, single-parent household in Los Angeles during the Great Depression, this "urban kid" (as she liked to refer to herself) was exposed early on to socioeconomic struggles; these formative experiences no doubt set the foundation for her contributions to social sciences and public policies later in life [90]. Her family conditions were so poor, they resorted to growing their own food in their backyard. She attended Beverly Hills High School. She believed it was her attendance at this high school that gave her the opportunity to go to college as "90% of kids who went to Beverly Hills High School went on to college." The fact that the high school was across the street from her home was serendipitous, as it seemed Ostrom would not have received support otherwise if she had not attended this high school. There are accounts of Ostrom mentioning how even her mother discouraged her from pursuing a college education, seeing no value in it; this only reinforces the early challenges Ostrom faced, even from her own family. However, despite the odds against her, Ostrom went on to attend and graduate from UCLA with honors [91].

Ostrom's journey into academia was further marked by challenges and gender biases. Despite her strong academic record, she was denied admission to UCLA's graduate economics program due to her lack of math background. Undeterred, she pursued a Ph.D. in political science, breaking through the barriers set by traditional gender norms. But some of Ostrom's retellings of her experience with prejudices is enough to send a shiver down anyone's spine:

> *"'I got circled in the schoolroom, out on the playground.'*
> *'You Jew! You Jew!' she recalled, her voice rising, imitating*
> *the taunts. 'Having that experience as a kid and being a woman,*
> *and having that challenge as it has been at different times to be*
> *a woman, I've got pretty good sympathy for people who are not*
> *necessarily at the center of civic appreciation."* [92]

These unfortunate experiences paved the way for the fierce character Ostrom would become, giving her a particular and illuminating perspective of life that shaped not only her existence but of those around her. She was an advocate for difference and plurality, promoting women and minorities, and her work fit nicely with feminist economics that challenged the traditionally male-biased field. Even though her research was mainly focused on commons and not around feminism and anti-racism, Ostrom's tireless efforts for equity were admirable [92].

Her marriage to Vincent Ostrom, a fellow political scientist, marked the beginning of a unique partnership in both life and academia. In 1965, they moved to Bloomington, Indiana, where Vincent accepted a position at Indiana University. Ostrom, however, faced gender discrimination and had to wait a year before securing a teaching position. Eventually, she became a crucial figure at the Workshop in Political Theory and Policy Analysis, a center they co-founded [91].

Ostrom's ground-breaking work focused on the commons and challenged established economic theories. She demonstrated that common-pool resources could be successfully managed by the people who use them, contradicting the prevailing notion that centralized regulation or privatization was necessary. Her research on understanding cases of failed and successful commons led to the development of ideas on overcoming "tragedy of the commons"; that under certain conditions, communities could self-organize and develop institutions to sustainably manage shared resources [93]. In her book *Governing the Commons* (1990), Ostrom advocated for communities to manage common-pool resources through collective action and self-governance. Ostrom used examples of past institutions in which such strategies proved to be successful; she identified the commonalities amongst these institutions and correlated this with their efficacy and survival [94].

Polycentricity, a concept she and her husband advocated, highlighted the importance of tailored, local solutions over centralized approaches. The Workshop they established was a testament to this philosophy, focusing on interdisciplinary collaboration and understanding real-world dynamics [92, 95].

Ostrom's achievements were not only a triumph over gender discrimination but also a testimony to her empirical approach. She believed in looking and listening, emphasizing the importance of understanding local circumstances and involving communities in designing and revising solutions. Her work highlighted the diversity of human communities and the need for bottom-up approaches to address their unique needs.

Elinor Ostrom's legacy extends beyond her Nobel Prize, as she continued her scholarly pursuits until her passing in 2012 [96]. Ostrom was referred to as someone who was comfortable in the field, and in every sense being a master of all. From working in the field to the laboratory, to running her own statistical and meta-analyses. "*Her research-methods quiver was full and she used every arrow in it*", remarked Shepsle (2010) [90]. Her commitment to understanding the real-world complexities of resource management and governance serves as an inspiration, challenging conventional wisdom and leaving an indelible mark on the field of economics.

# Lise Meitner

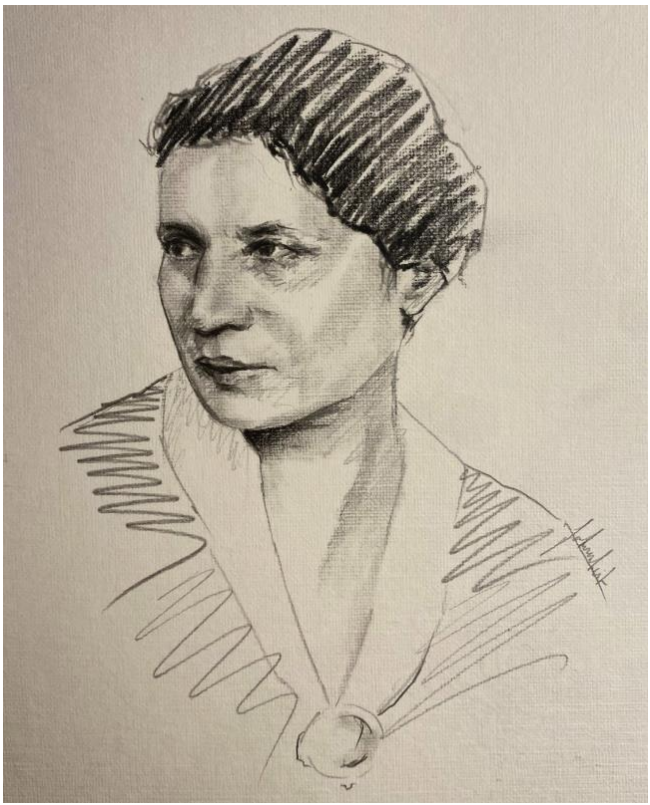

*"You must not blame us scientists for the use which war technicians have put our discoveries."* - Lise Meitner

**Figure 23. Lise Meitner.** Meitner was pivotal in the discovery of nuclear fission but shared a similar fate to many women of her time. The Nobel prize was instead awarded to colleague Otto Hahn. [*Illustrated by Farah Ajili, contributing author, 2024*]

Max Planck allegedly remarked that those born in the year 1879 were predestined to study physics. Amongst the ranks were Albert Einstein, Max von Laue and Otto Hahn. Lise Meitner (**Figure 23**) also should be included in this list. However, in comparison to other contemporary physicists of her era, Meitner experienced the harder life owing to her gender and religion [97]. Born in the heart of Vienna, amidst the intellectual fervor of the late 19th century, Meitner was born into a family that nurtured the seeds of knowledge. While raised in a Jewish household,

Lise (the third of eight children) was baptized as a Protestant. Her parents provided an intellectual environment that encouraged all children, regardless of gender, to pursue higher education, although it was still frowned upon for female students to pursue degrees in mathematics (thus why Meitner first focused on qualifying as a teacher of French) [97]. Later, in 1901, Lise entered the University of Vienna, a trailblazer among the first female students [98, 99].

Despite societal barriers (where Meitner herself would remark on the rudeness of fellow students, given a female student was regarded as a freak of nature), Lise excelled in her studies, delving into the realms of physics, chemistry, mathematics, and botany. In 1905, she achieved the highest honors in her doctoral thesis on the conduction of heat in inhomogeneous solids; she was the second woman in Vienna to achieve a doctorate in physics. Hindered by societal norms limiting women's academic careers, Lise's determination led her to Berlin in 1907, where she studied under Max Planck, a mentor who would play a pivotal role in her scientific journey [100].

Lise Meitner's resilience and brilliance soon caught the attention of chemist Otto Hahn – a man of the same age, and whose informal and frank nature appealed to the shy Meitner who was entirely comfortable with senior or authoritative figures. This collaboration lasted for three decades. There were, of course, challenges early on in their collaboration. While Meitner was happy to work with Hahn, his association with Emil Fischer (the supervisor Hahn was working under) made it difficult, given Fischer had a ban on women being in his laboratory at the time. This was overturned two years later when women's education became regularized and Fischer lifted the ban, but in the interim Meitner was relegated to working in a carpenter's workshop (or Fischer's "Holzwerkstatt") [100]. Facing challenges as a woman in a male-dominated field, Meitner and Hahn worked on various projects, including the discovery of isotopes and the element protactinium in 1918 [101]. However, despite their ground-breaking contributions, recognition eluded Meitner, as the academic world remained indifferent to the achievements of women.

In the midst of World War I, Lise shifted her focus to aiding the wounded as an X-ray equipment operator. After the war, she became the director of the physics department at her institute, continuing her collaboration with Hahn. The 1920s brought further triumphs as Lise unraveled the mysteries of radiation, discovering the Auger effect in 1923 (named after the French scientist Pierre Victor Auger, who discovered the phenomenon two years later) [100, 101].

As the political landscape shifted with the rise of Adolf Hitler, Meitner, a Jew protected by her Austrian nationality, clung to her position until 1938. With the Anschluss, she faced persecution and chose to flee to Sweden, leaving behind the tumultuous German regime. In Stockholm, Meitner persisted in her research, clandestinely collaborating with German scientists, including Otto Hahn, on the "uranium project."

In 1939, Lise's pivotal role in the discovery of nuclear fission emerged, revealing the potential for both scientific progress and destructive power. However, as the world grappled with the implications of this breakthrough, Lise found herself sidelined. The Nobel Prize in 1944 went to Otto Hahn alone, perpetuating the historical oversight of Lise Meitner's contributions.

Heisenberg stated: "*So far one would be inclined to regard the fission of uranium as the ultimate*

*triumph of many years of collaboration between the two scientists [Hahn and Meitner, although Frtiz Strassmann's name was omitted]."* During his lifetime, Hahn was reluctant to mention the name of Meitner in association with uranium fission. Strassmann additionally supported Meitner's stake in the discovery [97].

Despite the lack of formal recognition, Meitner continued her scientific journey, rejecting the opportunity to participate in the Manhattan Project and denouncing the creation of the atomic bomb. In 1968, she passed away in England, leaving behind a legacy that transcends the Nobel Prize. Element 109, meitnerium, was named in her honor in 1992 [102].

Lise Meitner, an indomitable force in the world of science, stands as a testament to the resilience and brilliance of women who, against all odds, have shaped the course of scientific history.

# Anatolian Women and the Daruşşifas (Hospitals)

While the theme of this article is on unheralded female scientists, we take a moment to give recognition to the Daruşşifas. Turkish dignitaries (high-ranking statesmen) and Atabegs (viziers), especially the sultan's wives and daughters, built madrasahs, soup kitchens, schools, roads, bridges, caravanserais, and hospitals everywhere. Foundations were allocated to cover their expenses. Hospitals (Daruşşifas) established by women were taken as examples in this study. These institutions, which were built to meet the health needs of the public, became popular thanks to foundations, and continued their duties for a long time. Physicians were trained in Hospitals (Daruşşifas), and physicians working in the surrounding Islamic and Turkish provinces worked in these organizations. Professional conversations were held between physicians and meetings were organized in here [103]. They were the first examples of modern health institutions.

**Gehver Nesibe Sultan Hospital (Daruşşifa)**

Turkish women established many institutions in Anatolia during the Middle Ages. These institutions were very advanced practices, considering the socio-cultural atmosphere of the period. Women at the time had limited opportunities and rights in both Europe and the Mediterranean Basin. There are important examples of Turkish women being active in social and cultural life. One of these women was Gevher Nesibe Hatun (**Figure 24**). As her will, Gevher Sultan requested that a healing center and an educational institution be founded in Kayseri [104]. Gevher Nesibe was founded as Çifteler or Gıyâsiye Hospital and Madrasa [105, 106]. According to the record of its inscription, the hospital was built in 1205 (602 h.) by her brother Gıyaseddin Keyhüsrev upon the will of Gevher Nesibe Hatun, the daughter of Sultan XI. Kılıç Arslan [107].

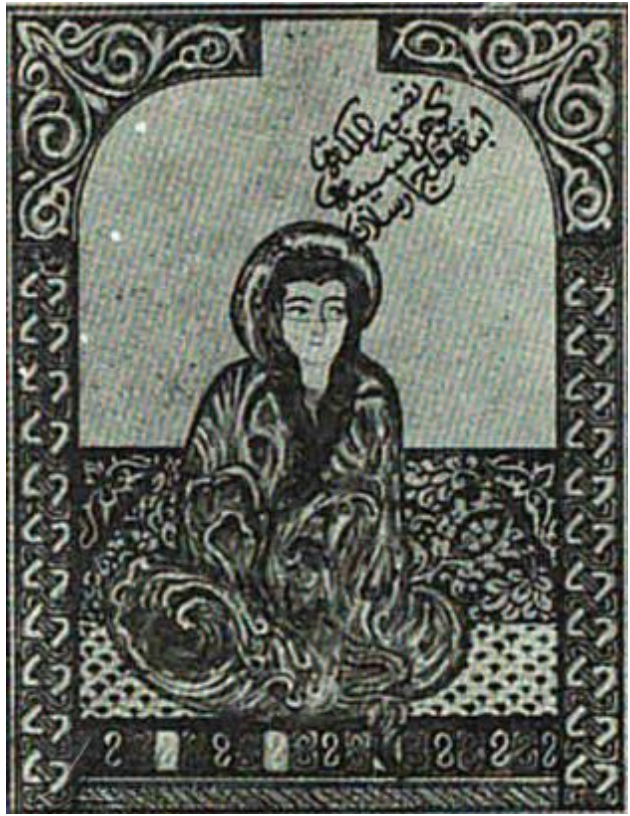

**Figure 24. Painting of Gevher Nesibe Sultan [108].** The role of Anatolian women in the creation of soup kitchens, schools, and most prominently hospitals (known as Daruşşifa in Arabic) is largely unknown. Gevher Nesibe Sultan founded the world's first independent Faculty of Medicine in 1205. The services provided by these women were instrumental in ensuring all members of society had access to basic human rights, such as healthcare, housing and food.

Gehver Nesibe Sultan (Maristan) Hospital (Daruşşifa), was opened in Kayseri in 1205, and served as both medical education, training and a hospital [109]. Bimaristan section was the place where clinical services were provided [110]. The world's first independent Faculty of Medicine was founded by Gevher Nesibe Hatun in Kayseri [111].

**Divriği Hospital (Darüşşifa)**

Divriği Hospital (Darüşşifa) was built by princess Turan Malik, daughter of Fahrettin Behram Şah from Menguceks. The translation of the inscription on the base of the hospital is as follows: (In order to gain the consent of the late ruler Fahrettin Behram Shah, the just Turan Sultan, who is in need of divine forgiveness, ordered the construction of this blessed (Ali) hospital in the year of Muharram 626 (December 1228) [112, 113].

Divriği Ulu Mosque and Hospital has four Gates [114]. Healing House Crown Temple, Mosque North Crown Gate, Mosque West Crown Gate and Shah Gathering Place Crown Gate. It is a dazzling marvel of architecture and engineering with unique decorations, each different from the other. It is described as the Hamra of Anatolia [115, 116].

**Amasya Hospital (Darüşşifa)**

It is located in the Yakutiye neighborhood of Amasya and is known as the Tımarhane. It is a marvel of art whose doors are decorated with floral decorations. The following information is given on the Hospital's door base. It is learned that the Hospital was built in 708 (1308-1309) by Anber Ağa. Abdullah and the Anatolian Emir Ahmed Bek, under the auspices of İlduş (Yıldız) Hatun, the wife of the Ilkhanid Ruler Sultan Olcaytu Mehmed Khan [117]. It was also used for a long time as a place where mental patients were treated [118]. Sabuncuoğlu Şerefeddin, a physician and surgeon from Amasya, conducted clinical research and practice at the Hospital (Darüşşifa) for fourteen years [119]. Today, it is known as Sabuncuoğlu Şerefeddin Hospital.

According to the inscriptions, it is seen that these historical buildings were a hospital. On the other hand, the names of girls belonging to the Seljuk family were generally not expressed clearly in this way. As a general expression, it was used as "İsmeti'd-dunya ve'd-din". In the inscriptions, direct references are made to women, as in Gevher Nesibe Hatun. In this case, just as Turkish women were active in social life, they were also actively involved in such scientific and artistic activities. However, their names are not well known.

It reveals that in Seljuk hospitals, patients were treated without discrimination of rich, poor, religion, language, and race. In these hospitals, medicines and food were given to patients free of charge, and their treatments were also provided free of charge [120]. In these three sample institutions, special rooms were allocated for mental illnesses and music therapy was applied among the treatment methods. At the same time, special diet lists were prepared according to the patients.

# The Next Generation: Farida Fassi

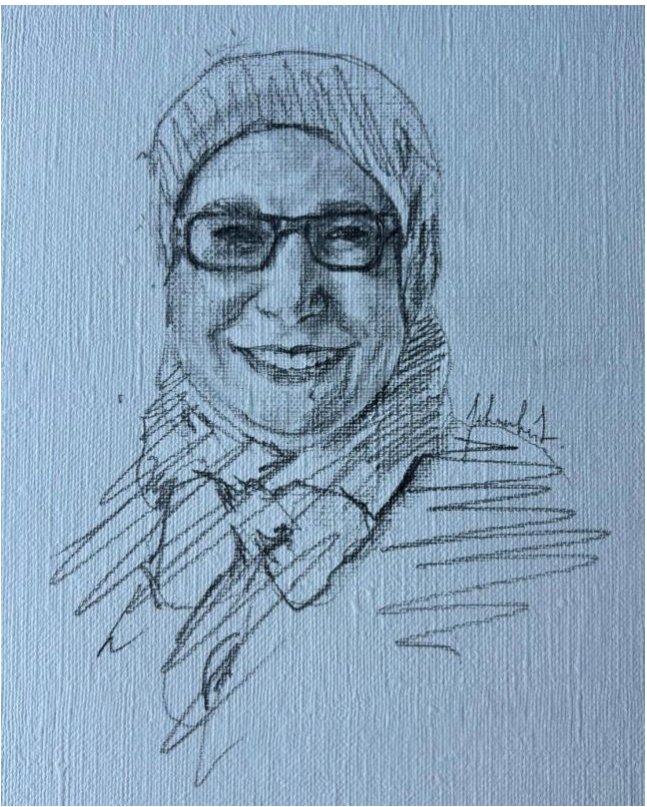

**Figure 25. Farida Fassi.** A professor of physics, Fassi was ranked among the top 50 scientists worldwide. [*Illustrated by Farah Ajili, contributing author, 2024*]

*"You are smart enough and capable enough to join any field in STEM, and you are equally as deserving to be in these fields as anybody else. Follow your curiosities and believe in yourself"* –
Farida Fassi

Farida Fassi (**Figure 25**) is a Moroccan researcher and a professor of physics that was ranked among the 50 top scientists worldwide according to the AD Scientific Index in 2021 [121]. She began her academic journey by earning a Bachelor in Physical Sciences in Morocco before moving to Spain to further her studies. There, she completed her Masters and Ph.D. in Particle Physics, notably participating in the ATLAS collaboration at CERN, aimed at researching the Higgs boson, a fundamental particle in understanding the origin of matter. This experience enabled her to contribute to the construction and calibration of the ATLAS detector, a complex device used to identify and study elementary particles created during proton collisions within the Large Hadron Collider (LHC).

After defending her thesis, Farida Fassi continued her career in France, where she held a postdoctoral position in Lyon, working on data control and processing for the LHC's CMS experiment. She also continued her research in particle physics, focusing on the precise measurement of the top quark.

Throughout her journey, Farida Fassi faced challenges, but her commitment to science and her desire to contribute to her country's development led her back to Morocco after nearly two decades in Europe. Despite the obstacles encountered in the Moroccan scientific community, particularly in terms of research organization and funding, Farida Fassi remains determined to share her expertise and encourage young Moroccan scientists. Currently a professor at Mohammed V University in Rabat, Farida Fassi continues to conduct research in particle physics and collaborates with international institutions to promote science development in Morocco. She is also active in organizing conferences and workshops for doctoral students, aiming to prepare them for successful careers in scientific research. Despite the challenges, Farida Fassi remains optimistic and determined to contribute to scientific progress and inspire the next generation of researchers in Morocco.

Her exemplary journey demonstrates her passion for physics and her commitment to advancing scientific knowledge in her country and internationally. Farida Fassi was also the first veiled woman to work at CERN. Her experience at CERN allowed her to collaborate with researchers from around the world and make significant contributions to particle physics research. As a trailblazer, she paved the way for other women from diverse cultural and religious backgrounds, demonstrating that science is accessible to all, regardless of gender, ethnic origin, or religion. Farida Fassi's journey not only reflects her dedication to science but also her ability to overcome obstacles and break stereotypes. As a role model and mentor, she inspires girls and women worldwide to pursue their scientific dreams, regardless of the barriers they may face. Her story underscores the importance of diversity and inclusion in scientific research and highlights the transformative potential of passion and perseverance.

## Final Remarks

It was difficult to do justice to the remarkable individuals described in this paper in a concise and insightful manner. As we conducted our research into their lives, their heartbreaks and great achievements, we felt connected and overwhelmed by their struggles and eventual triumphs. We began to understand their pain, their anger, and the burden of injustice that they suffered. Many of these women passed away before gaining recognition for their great achievements. We realised that we were not simply writing a piece that depicted the past, but a lesson for guiding the future of scientific research, where such trials and tribulations are relegated to the pages of history.

# Author Information

## Authors and Affiliations

**Sorbonne Université, Institut du Cerveau—Paris Brain Institute—ICM, CNRS, Inria, Inserm, AP-HP, Hopital de la Pitié Salpetriere, Paris, France**

Janan Arslan, Lina Sami, Daniela Domingues & Violetta Zujovic

**Technology, Policy and Management, Delft University of Technology, Delft, The Netherlands**

Sepinoud Azimi & Lærke Vinther Christiansen

**Maison des Artistes, Montpellier, France**

Farah Ajili

**Erzincan Binali Yıldırım University, Faculty of Education Department of Social Sciences Education, Erzincan, Turkiye**

Deniz Akpinar

**School of Engineering, University of Melbourne, Parkville, Victoria, Australia**

Kurt K. Benke

# Contributions

The research was designed, and the analysis was performed by J.A. Authors J.A., S.A., L.S., D.D., D.A., L.V.C., V.Z., & K.K.B. have contributed equally to the writing of the manuscript. Author F.A. contributed to all the portraits. Authorship order simply represents the sequence in which members joined the project. All authors have the right to list their name first in their respective profiles and portfolios.

# Corresponding author

Correspondence to Janan Arslan ([janan.arslan@icm-institute.org](janan.arslan@icm-institute.org)).

## Code availability

Code created using R to create the timeseries plots and choropleth maps are provided with this paper in **Supplementary Information**.

## Data availability

Data collected using Google Trends are provided with this paper in **Supplementary Information**.

# References

1. Gage, M.J., *Woman as an Inventor.* The North American Review, 1883. **136**(318): p. 478-489.
2. Rossiter, M.W., *The Matthew Matilda Effect in Science.* Social Studies of Science, 1993. **23**(2): p. 325-341.
3. Merton, R.K., *The Matthew Effect in Science.* Science, 1968. **159**(3810): p. 56-63.
4. Lincoln, A.E., et al., *The Matilda Effect in science: Awards and prizes in the US, 1990s and 2000s.* Social Studies of Science, 2012. **42**(2): p. 307-320.
5. Patel, S.R., et al., *The Matilda Effect: Underrecognition of Women in Hematology and Oncology Awards.* The Oncologist, 2021. **26**(9): p. 779-786.
6. Rapoport, S., *Rosalind Franklin: Unsung Hero of the DNA Revolution.* New York History, 2003. **84**(3): p. 315-329.
7. Pebesma, E. and R. Bivand, *Spatial Data Science: With Applications in R*. 1st ed. 2023: Chapman and Hall/CRC.
8. Wickham, H., *ggplot2: Elegant Graphics for Data Analysis*. 2016, New York: Springer-Verlag.
9. Massicotte, P. and A. South, *rnaturalearth: World Map Data from Natural Earth*. 2023.
10. *Rosalind Franklin, DNA scientist, celebrated by Google doodle*. 2013 28/07/2024]; Available from: https://www.theguardian.com/technology/2013/jul/25/rosalind-franklin-google-doodle.
11. Parkinson, C. and I. Labrouillere, *A Critical Companion to Christopher Nolan*. 2023: Rowman & Littlefield.
12. Shearer, S.M., *Glamour and Style: The Beauty of Hedy Lamarr*. 2022: Rowman & Littlefield.
13. Shearer, S.M., *Beautiful: The Life of Hedy Lamarr*. 2010, New York: Thomas Dunne Books/St. Martin's Press.
14. Thomas, L., *Actress Hedy Lamarr, Inventor: A Public Image Reframed*, in *Interdisciplinary Centre for Culture and Creativity*. 2022, University of Saskatchewan.
15. Wallmark, L. and K. Wu, *Hedy Lamarr's Double Life: Hollywood Legend and Brilliant Inventor*. 2019, New York: Sterling Children's Books.
16. Barton, R., *Hedy Lamarr: The Most Beautiful Woman in Film*. 2010: University Press of Kentucky.
17. Rosenberg, M. and S.D. Picker, *Historical Heroines : One Hundred Women You Should Know About*. 2018, Havertown, UNITED STATES: Pen & Sword Books Limited.
18. *George Antheil.* The Musical Times, 1959. **100**(1394): p. 220-220.
19. Abelson, H., K. Ledeen, and H. Lewis, *Blown to Bits: Your Life, Liberty, and Happiness After the Digital Explosion*. 2008, Crawfordsville, Indiana: Addison-Wesley.

20. Pellinen-Wannberg, A., *Women in radio science: Introduction by the associate editor.* URSI Radio Science Bulletin, 2020. **2020**(372): p. 61-62.
21. Markström, K.A., *The invention by Hedy Lamarr and George Antheil of frequency-hopping spread-spectrum secret communications.* URSI Radio Science Bulletin, 2020. **2020**(372): p. 62-63.
22. Lejeune, J., M. Gauthier, and R. Turpin, *[Human chromosomes in tissue cultures].* C R Hebd Seances Acad Sci, 1959. **248**(4): p. 602-3.
23. Lejeune, J., M. Gautier, and R. Turpin, *THE CHROMOSOMES OF MAN.* The Lancet, 1959. **273**(7078): p. 885-886.
24. Gautier, M. and P.S. Harper, *Fiftieth anniversary of trisomy 21: returning to a discovery.* Human Genetics, 2009. **126**(2): p. 317-324.
25. Fleming, N., *J'ACCUSE.* New Scientist, 2014. **221**(2963): p. 44-47.
26. Casassus, B., *Down's syndrome discovery dispute resurfaces in France.* Nature, 2014.
27. L'Inserm, M.C.d.É.d., *Avis du Comité d'éthique de l'Inserm relatif à la saisine d'un collectif de chercheurs concernant la contribution de Marthe Gautier dans la découverte de la trisomie 21*. 2014.
28. *Marthe Gautier*. 2015 [cited 2024 23/01/2024]; Available from: https://www.inserm.fr/portrait/histoire/marthe-gautier/.
29. Pain, E. *After More Than 50 Years, a Dispute Over Down Syndrome Discovery*. 2014 [cited 2024 23/01/2024]; Available from: https://www.science.org/content/article/after-more-50-years-dispute-over-down-syndrome-discovery.
30. *Jérome Lejeune pioneer in the discovery of Trisomy 21*. 2018 [cited 2024 23/01/2024]; Available from: https://www.fondationlejeune.org/assets/uploads/2022/06/DISCOVER-OF-DOWN-SYNDROME_ENGLISH_Sept.2018.pdf.
31. Dorothy, M.C.H., *Kathleen Lonsdale. 28 January 1903 -- 1 April 1971.* Biographical Memoirs of Fellows of the Royal Society, 1975. **21**: p. 447-484.
32. Wilson, J.M., *Dame Kathleen Lonsdale FRS (1903–1971): her contribution to crystallography.* ChemTexts, 2021. **7**(4): p. 23.
33. Lonsdale, K. and R. Whiddington, *The structure of the benzene ring in $C_6$ $(CH_3)_6$.* Proceedings of the Royal Society of London. Series A, Containing Papers of a Mathematical and Physical Character, 1929. **123**(792): p. 494-515.
34. Lonsdale, K., *X-ray evidence on the structure of the benzene nucleus.* Transactions of the Faraday Society, 1929. **25**: p. 352-366.
35. *Admission of women into the Fellowship of the Royal Society.* Notes and Records of the Royal Society of London, 1946. **4**(1): p. 39-40.
36. *KATHLEEN LONSDALE,* in *These Strange Criminals*, P. Brock, Editor. 2004, University of Toronto Press. p. 232-242.
37. Baldwin, M., *'WHERE ARE YOUR INTELLIGENT MOTHERS TO COME FROM?': MARRIAGE AND FAMILY IN THE SCIENTIFIC CAREER OF DAME KATHLEEN*

*LONSDALE FRS (1903-71)*. Notes and Records of the Royal Society of London, 2009. **63**(1): p. 81-94.
38. Rayner-Canham, M. and G. Rayner-Canham, *Marguerite Perey: The Discoverer of Francium*, in *Women in Their Element*. 2019, WORLD SCIENTIFIC. p. 341-349.
39. Scerri, E.R., *Master of Missing Elements*. American Scientist, 2014. **102**(5): p. 358-365.
40. Draganić, I., Z.D. Draganić, and J.-P. Adloff, *Radiation and radioactivity on earth and beyond*. 2020, Boca Raton: CRC Press.
41. Weaver, E.C., *Names for Chemical Elements*. The Science Teacher, 1949. **16**(4): p. 184-184.
42. Scerri, E., *Finding francium*. Nature Chemistry, 2009. **1**(8): p. 670-670.
43. Adloff, J.-P. and G.B. Kauffman, *Francium (Atomic Number 87), the last discovered natural element*. Chem. Educ, 2005. **10**(5): p. 387-394.
44. Leridon, H., *The Demography of a Learned Society: The Académie des Sciences (Institut de France), 1666-2030*. Population (English Edition, 2002-), 2004. **59**(1): p. 81-114.
45. Mason, J., *The Women Fellows' Jubilee*. Notes and Records of the Royal Society of London, 1995. **49**(1): p. 125-140.
46. *Vital Statistics*. The British Medical Journal, 1962. **1**(5280): p. 809-812.
47. Adloff, J.-P. and G.B. Kauffman, *Triumph over prejudice: the election of radiochemist Marguerite Perey (1909–1975) to the French Académie des Sciences*. The Chemical Educator, 2005. **10**: p. 395-399.
48. Preston, S.S., *Marguerite Perey (1909–1975): Discoverer of Francium*, in *The Posthumous Nobel Prize in Chemistry. Volume 2. Ladies in Waiting for the Nobel Prize*. 2018, ACS Publications. p. 245-263.
49. HOLLMANN, H.T., *THE FATTY ACIDS OF CHAULMOOGRA OIL IN THE TREATMENT OF LEPROSY AND OTHER DISEASES*. Archives of Dermatology and Syphilology, 1922. **5**(1): p. 94-101.
50. Mushtaq, S. and P. Wermager, *Alice Augusta Ball: The African-American chemist who pioneered the first viable treatment for Hansen's Disease*. Clin Dermatol, 2023. **41**(1): p. 147-158.
51. Wermager, P. and C. Heltzel, *Alice A. Augusta Ball Young Chemist Gave Hope to Millions*. ChemMatters, 2007. **16**.
52. Cederlind, E., *A tribute to Alice Bell: a scientist whose work with leprosy was overshadowed by a white successor*. The Daily of the University of Washington, 2008. **29**(02).
53. McWilliams Tullberg, R., *Women at Cambridge*. 1998, United Kingdom: Cambridge University Press.
54. Wayman, P.A., *Cecilia payne-gaposchkin: astronomer extraordinaire*. Astronomy & Geophysics, 2002. **43**(1): p. 1.27-1.29.
55. Russell, H.N., *On the composition of the Sun's atmosphere*. Astrophysical Journal, vol. 70, p. 11, 1929. **70**: p. 11.
56. Moore, D., *What stars are made of: The life of Cecilia Payne-Gaposchkin*. 2020: Harvard University Press.

57. Payne-Gaposchkin, C., *The stars of high luminosity.* New York and London, Pub. for the Observatory by the McGraw-Hill book company, inc., 1930., 1930. **3**.
58. Weyl, H., *Emmy Noether.* Scripta Mathematica, 1935. **3**(3): p. 201-220.
59. Lederman, L.M. and C.T. Hill, *Symmetry and the beautiful universe.* 2011: Prometheus books.
60. Gilmer, R., *Commutative ring theory.* 1981.
61. Kimberling, C., *Emmy Noether and her influence.* Emmy Noether: A tribute to her life and work, 1981: p. 1-64.
62. Dick, A. and H. Weyl, *Emmy Noether*. 1981: Springer.
63. Blomberg, G., *Flora Tristan: a predecessor of Marx and Engels.* EDITOR: Erwin Marquit (physics, Univ. of Minnesota) MANUSCRIPT EDITOR: Leo Auerbach (English education, retired, Jersey City State College) EDITORIAL STAFF: Gerald M. Erickson, April Ane Knutson, Doris, 1998. **11**(1): p. 5.
64. Cross, M.F., *The relationship between feminism and socialism in the life and work of Flora Tristan (1803-1844)*. 1988, Newcastle University.
65. Grogan, S.K. and S.K. Grogan, *Flora Tristan and the Moral Superiority of Women.* French Socialism and Sexual Difference: Women and the New Society, 1803–44, 1992: p. 155-174.
66. Tristan, F., *The Workers' Union (B. Livingston, Trans.).* . 1983: University of Illinois Press.
67. Talbot, M., *An Emancipated Voice: Flora Tristan and Utopian Allegory.* Feminist Studies, 1991. **17**(2): p. 219-239.
68. Walsh, L. *Journey of Discovery*. 04/02/2024]; Available from: https://www.cam.ac.uk/stories/journeysofdiscovery-pulsars.
69. Tesh, S. and J. Wade, *'Look happy dear, you've just made a discovery'.* Physics World, 2017. **30**(9): p. 31.
70. *Avenue Interviews*. 2023 04/02/2024]; Available from: https://www.gla.ac.uk/explore/avenue/me/mebyjocelynbellburnell/.
71. Hewish, A., et al., *Observation of a Rapidly Pulsating Radio Source.* Nature, 1968. **217**(5130): p. 709-713.
72. , Y.S.J. Ailís, Editor., Young Scientists Journal.
73. Geleff, K., C. Howden, and K. Ball. *He got the Nobel. She got nothing. Now she's won a huge prize and she's giving it all away*. 2018 04/02/2024]; Available from: https://www.cbc.ca/radio/asithappens/as-it-happens-monday-edition-1.4817159/he-got-the-nobel-she-got-nothing-now-she-s-won-a-huge-prize-and-she-s-giving-it-all-away-1.4817161.
74. *Nobel Prize Website*. 1974 [cited 04/02/2024; Available from: https://www.nobelprize.org/prizes/physics/1974/summary/.
75. Jaeger, L., *Women of Genius in Science: Whose Frequently Overlooked Contributions Changed the World*. 2023: Springer Nature.
76. Physics, I.o. *Jocelyn Bell Burnell: the woman behind the fund*. 04/02/2024]; Available from: https://www.iop.org/about/support-grants/bell-burnell-fund/woman-behind-fund.
77. Roth, G., *Marianne Weber and her circle.* Society, 1990. **27**(2): p. 63-69.

78. Wobbe, T., *Elective affinities: Georg Simmel and Marianne Weber on gender and modernity.* Engendering the Social: Feminist Encounters with Sociological Theory, 2004: p. 54-68.
79. Becker, H. and M. Weber, *Max Weber, Assassination, and German Guilt: An Interview with Marianne Weber.* The American Journal of Economics and Sociology, 1951. **10**(4): p. 401-405.
80. Ulrich, L.T., *Vertuous Women Found: New England Ministerial Literature, 1668-1735.* American Quarterly, 1976. **28**(1): p. 20-40.
81. Lengermann, P.M. and G. Niebrugge, *The women founders: Sociology and social theory 1830–1930, a Text/Reader*. 2006: Waveland Press.
82. Hanke, E., *"'Max Weber's Desk is now my Altar': Marianne Weber and the intellectual heritage of her husband".* History of European Ideas, 2009. **35**(3): p. 349-359.
83. Chiang, T.-C., *Madame Wu Chien-Shiung: The first lady of physics research*. 2013: World Scientific.
84. Wu, C.-S., *Chien-Shiung Wu.* Physicists, 2014: p. 165.
85. Indumathi, D., *Chien-Shiung Wu: The First Lady of Physics.* Resonance, 2020. **25**(3): p. 333-352.
86. Reed, B.C., *The history and science of the Manhattan Project*. 2014: Springer.
87. Lau, C.D., K. Chow, and A.A. Ejaz, *Chien-Shiung Wu, ºft#(1912–1997) 1978 Wolf Laureate in Physics.* Nobel And Lasker Laureates Of Chinese Descent: In Literature And Science, 2019: p. 245.
88. Calvin, S., *Beyond Curie: Four women in physics and their remarkable discoveries, 1903 to 1963*. 2017: Morgan & Claypool Publishers.
89. Benczer-Koller, N., *CS W (1912-1997).* Biographical Memoirs National Academy of Sciences, 2009. **2009**: p. 1.
90. Shepsle, K.A., *Elinor Ostrom: uncommon.* Public Choice, 2010. **143**(3/4): p. 335-337.
91. Tarko, V., *Elinor Ostrom : An Intellectual Biography*. 2016, Blue Ridge Summit, UNITED KINGDOM: Rowman & Littlefield Publishers, Incorporated.
92. Wall, D., *Feminism and Intersectionality*, in *Elinor Ostrom's Rules for Radicals*. 2017, Pluto Press. p. 69-78.
93. Wall, D., *The Commons From Tragedy to Triumph*, in *Elinor Ostrom's Rules for Radicals*. 2017, Pluto Press. p. 21-34.
94. Bergstrom, T.C., *The Uncommon Insight of Elinor Ostrom.* The Scandinavian Journal of Economics, 2010. **112**(2): p. 245-261.
95. Cole, D.H., et al., *Elinor Ostrom and the Bloomington School of Political Economy : Polycentricity in Public Administration and Political Science*. 2014, Blue Ridge Summit, UNITED STATES: Lexington Books/Fortress Academic.
96. Herzberg, R. and B. Allen, *OBITUARY: Elinor Ostrom (1933—2012).* Public Choice, 2012. **153**(3/4): p. 263-268.
97. Krafft, F., *Lise Meitner: Her Life and Times—On the Centenary of the Great Scientist's Birth.* Angewandte Chemie International Edition in English, 1978. **17**(11): p. 826-842.

98. Sime, R.L., *Lise Meitner: a 20th century life in physics*. Endeavour, 2002. **26**(1): p. 27-31.
99. Sime, R.L., *Lise Meitner: A life in physics*. 1996: Univ of California Press.
100. Frisch, O.R., *Lise Meitner, 1878-1968*. 1970, The Royal Society London.
101. Sime, R.L., *Lise Meitner's escape from Germany.* American Journal of Physics, 1990. **58**(3): p. 262-267.
102. Abbott, S., *Lise Meitner and the dawn of the nuclear age, by Patricia Rife. Pp. 432. SFr68. 1998. ISBN 3 7643 3732 X (Birkhäuser)*. The Mathematical Gazette, 1999. **83**(498): p. 538-539.
103. Ünver, A.S., *Selçuklu Tababeti*. 2014, Ankara: Türk Tarih Kurumu.
104. Köker, A.H., *Gevher Nesibe Dârüşşifâsı Ve Tıp Medresesi*. 1992, Istanbul: Tıbbî Etik Yıllığı, II.
105. Ethem, H., *Selçukî tarihinden bir kıt'a, Tarih-i Osmanî Encümeni*. 1918, Kayseriye şehri mebâni-i İslâmiye ve kitabeleri: Kayseri.
106. Süheyl, A., *İlk Tip Mektebi ve Seririyatımız Sekiz Asır Evvel Kayseri'de Açılmıştır. 602h-1205m.* . Tedavi Seririyatı ve Laboratuvarı, 1932. **5**(2): p. 46-53.
107. Yinanç, R., *Kayseri ve Sivas Darüşşifaları'nın Vakıfları*. Belleten, 1984. **48**(189-190): p. 299-308.
108. Ünver, A.S. *Painting of Gevher Nesibe Sultan*. [cited 2024; Available from: https://portal.yek.gov.tr/works/detail/410750.
109. Akpınar, D. and E. Erhan, *Türkiye'de Bakteriyoloji Alanında Gelişmeler (1876-1938)*. 2020: Kutlu Yayınevi.
110. Inan, A., *Türk Kadınlarının Yaptırdıkları Sağlık Kurulları ve Gevher Nesibe Şifaiyyesi*. Tüberküloz ve Toraks Tıbbi Mecmua, 1955. **3**(2): p. 138-145.
111. Ünver, A.S., *Kayseri Tıp Sitemiz 760 Yaşında*. 1966, Istanbul: İsmail Akgün Matbaası, .
112. Süheyl, A., *Divrikide Prenses Turan Melik Hastanesi 1228*. Tedavi Seririyatı ve Laboratuvarı, 1934. **15**(4): p. 126-129.
113. Ülgen, A.S., *Divriği Ulu Camii ve Darüşşifası*. Vakıflar Dergisi, V., 1962: p. 94.
114. Huart, C., *Divriği Ulu Camii ve Darüşşifası Kuruluşunun 750. Yılı Hatıra Kitabı*. Journal Asiatique, 1978: p. 413.
115. Çelebi, E., *Seyahatname (Türkçeleştiren: Zuhuri Danışman)*. 1970, C. V, İstanbul.
116. Ünver, A.S., *Anadolu Selçuklularında Sağlık Hizmetleri*. 1972, Ankara: Malazgirt Armağanı.
117. Hüsameddin, H., *Amasya Tarihi*. 1327. p. 171-175.
118. Hüsameddin , A.A., *Amasya Tarihi*. 1328, Hikmet Matbaası: Dersâdet. p. 171.
119. Sabuncuoğlu, Ş.b.A.b.İ.e., *Cerrâhiyyetü'l-Hâniyye*.
120. Acıduman , A., *Darüşşifalar Bağlamında Kitabeler, Vakıf Kayıtları ve Tıp Tarihi Açısından Önemleri - Anadolu Selçuklu Darüşşifaları Özelinde*, A.Ü.T.F. Mecmuası, Editor. 2010. p. 10.
121. Kasraoui, S. *Two Moroccan Women Scientists Feature in AD Scientific Index 2023*. 2022 12/05/2024]; Available from: https://www.moroccoworldnews.com/2022/10/351796/two-moroccan-women-scientists-feature-in-ad-scientific-index-2023.